\documentclass[10pt,journal,compsoc]{IEEEtran}
\ifCLASSOPTIONcompsoc
  \usepackage[nocompress]{cite}
\else
  \usepackage{cite}
\fi
\IEEEoverridecommandlockouts
\usepackage{amsmath,amssymb,amsfonts}
\usepackage{algorithmic}
\usepackage{graphicx}
\usepackage{textcomp}
\usepackage{xcolor}
\def\BibTeX{{\rm B\kern-.05em{\sc i\kern-.025em b}\kern-.08em
    T\kern-.1667em\lower.7ex\hbox{E}\kern-.125emX}}
\usepackage{graphicx}
\usepackage{subcaption}
\usepackage[utf8]{inputenc}
\usepackage{fancyhdr}
\usepackage{dirtytalk}
\usepackage{tabularx}
\PassOptionsToPackage{hyphens}{url}\usepackage{hyperref}
\usepackage[ruled,lined,linesnumbered]{algorithm2e}

\usepackage{enumitem,kantlipsum}

\usepackage{graphicx}
\usepackage[utf8]{inputenc}
\usepackage{fancyhdr}
\usepackage{dirtytalk}
\usepackage{tikz}
\usetikzlibrary{shapes.geometric, arrows}
\usepackage{pgfplots}
\pgfplotsset{width=8cm,compat=1.16}
\usepgfplotslibrary{external}
\usepackage{mathtools}

\newtheorem{theorem}{\normalfont{\textbf{Theorem}}}

\DeclareMathOperator*{\argmax}{argmax}

\tikzstyle{startstop} = [rectangle, rounded corners, minimum width=2cm, minimum height=0.5cm,text centered, draw=black, fill=red!30]
\tikzstyle{process} = [rectangle, minimum width=2cm, minimum height=0.5cm, text centered, draw=black, fill=orange!30, align=left]
\tikzstyle{decision} = [diamond, minimum width=1.0cm, minimum height=0.4cm, text centered, draw=black, fill=green!30]
\tikzstyle{arrow} = [thick,->,>=stealth]

\newcolumntype{L}{>{$}l<{$}}

\usepackage[labelformat=simple]{subcaption}

\DeclareMathOperator*{\argmin}{arg\,min}
\newcolumntype{P}[1]{>{\centering\arraybackslash}p{#1}}

\begin{document}
\title{On The Robustness of Channel Allocation in Joint Radar And Communication Systems:\\ An Auction Approach}

\author{Ismail~Lotfi, Hongyang~Du, Dusit~Niyato~\IEEEmembership{,~Fellow,~IEEE}, Sumei~Sun~\IEEEmembership{,~Fellow,~IEEE}, and Dong~In~Kim~\IEEEmembership{,~Fellow,~IEEE}
 \IEEEcompsocitemizethanks{\IEEEcompsocthanksitem Ismail~Lotfi, Hongyang~Du and Dusit~Niyato are with the School of Computer Science and Engineering, Nanyang Technological University, Singapore 639798, Singapore\\
 E-mail: ISMAIL003@e.ntu.edu.sg; HONGYANG001@e.ntu.edu.sg; DNIYATO@ntu.edu.sg.
 \IEEEcompsocthanksitem Sumei Sun is with the Institute for Infocomm Research, A*STAR, Singapore 138632, Singapore\\
 E-mail: sunsm@i2r.a-star.edu.sg.
 \IEEEcompsocthanksitem Dong In Kim is with the Department of Electrical and Computer Engineering, Sungkyunkwan University (SKKU), Suwon 16419, South Korea\\ E-mail: dikim@skku.ac.kr.
 }
 }

\IEEEtitleabstractindextext{\begin{abstract}
Joint radar and communication (JRC) is a promising technique for spectrum re-utilization, which enables radar sensing and data transmission to operate on the same frequencies and the same devices. However, due to the multi-objective property of JRC systems, channel allocation to JRC nodes should be carefully designed to maximize system performance. Additionally, because of the broadcast nature of wireless signals, a watchful adversary, i.e., a warden, can detect ongoing transmissions and attack the system. Thus, we develop a covert JRC system that minimizes the detection probability by wardens, in which friendly jammers are deployed to improve the covertness of the JRC nodes during radar sensing and data transmission operations. Furthermore, we propose a robust multi-item auction design for channel allocation for such a JRC system that considers the uncertainty in bids. The proposed auction mechanism achieves the properties of truthfulness, individual rationality, budget feasibility, and computational efficiency. The simulations clearly show the benefits of our design to support covert JRC systems and to provide incentive to the JRC nodes in obtaining spectrum, in which the auction-based channel allocation mechanism is robust against perturbations in the bids, which is highly effective for JRC nodes working in uncertain environments.
\end{abstract}


}
\maketitle
\IEEEdisplaynontitleabstractindextext
\IEEEpeerreviewmaketitle

\IEEEraisesectionheading{\section{Introduction}}
\IEEEPARstart{W}{ith} the growing number of mobile users in cellular networks and given the limited available bandwidth, several approaches have been proposed at different levels of the network design to accommodate these users and efficiently allocate the spectrum. 
In addition, different technologies are being integrated into these network devices to perform a variety of tasks such as communication, radar sensing, and power transfer~\cite{Liu2020_survey, Ismail_2021_TVT}. This allows efficient spectrum re-utilization and minimizes hardware costs. Recently, spectrum sharing between radar and communication systems has led to the emergence of joint radar and communication (JRC) systems~\cite{Liu2020_survey}. In JRC systems, simultaneous radar sensing and data transmission can be performed by the same device on the same frequency. However, JRC systems are vulnerable to jamming and eavesdropping attacks, and only a few works address this issue. 

In our previous work~\cite{Ismail_2021_TVT}, we have developed a framework for JRC systems that enables continuous operation of the system under jamming attacks. We developed an intelligent deception strategy based on ambient backscatter communication and deep reinforcement learning. However, the proposed model faces some scalability issues. First, if the network has multiple JRC nodes that are communicating with a receiver, it becomes harder for the receiver to decode the modulated ambient backscatter signals. Second, a warden can still detect ongoing radar sensing signals and knows exactly which JRC node is trying to sense the environment, hence making the JRC node potentially vulnerable to attackers.
The warden can eavesdrop the transmitted signals (communication signals and radar signals) to infer sensitive information about the users. Therefore, minimizing the detection probability by a warden is crucial for preventing such attacks as the warden will be unable to distinguish between noise and real ongoing transmission and sensing activities.
As such, hiding the transmitted signals can both minimize the leakage of sensitive information and prevent jamming attacks from occurrence.

Recently, covert communication has been proposed as an effective solution to hide the ongoing signal transmission from discovery by a third party~\cite{Bash_2013, Jiang_2021}. Specifically, the idea consists of deploying friendly jammers to transmit fake signals to disturb the warden about the ongoing transmission. Since the warden has a predefined energy threshold to detect ongoing communication, the addition of jamming signals will increase the false alarm and misdetection probabilities at the warden. As a result, the warden is forced to increase its detection threshold and allows the legitimate transmitter to transmit silently and covertly thanks to the jamming signals.

Since the covert communication technique can be applicable and compatible with any wireless systems without changing the system operation, it is a low-cost and efficient solution, for example, compared with the frequency hopping (FH) technique which requires the transmitters and receivers to switch among subcarriers frequently. As such, the FH technique increases the complexity to the circuit design and requires a sophisticated synchronization method. Moreover, the warden could scan a wide range of frequencies, overcoming the FH technique. Therefore, covert communications can avoid other external attacks as the transmission is hidden. \textcolor{black}{With a variety of modulation schemes developed for JRC systems such as orthogonal frequency division multiplexing (OFDM)~\cite{Liu2017_OFDM, Sturm_2011}, the idea of covert communication can be extended to enable covert radar sensing and hence, a covert JRC system. 
However, radio resource management issues arise when there are multiple JRC nodes in the network. Specifically, these JRC nodes can have different spectrum requirements to achieve their targeted data rates and sensing operations. 
The JRC nodes need to carefully design their valuation functions based on their private information, e.g., channel state information (CSI) and importance of radar sensing over communication and vice-versa, to ensure that the obtained spectrum from the SSP is worth its price, i.e., positive utility.}

\textcolor{black}{
Even though several works have studied at the signal processing level a coexistence of radar sensing and data transmission on the same waves and/or on the same devices~\cite{Liu2020_survey}, optimal channel allocation for JRC nodes has received far less attention. 
In particular, a major part of existing works focused on finding the optimal scheduling and modulation scheme for radar and communication signals, see for example the comprehensive review in~\cite{Luong_2021_JRC_survey}. However, fewer works addressed the problem of allocating the available bandwidth to the JRC nodes so that the utilities of both the SSP and the JRC nodes are maximized~\cite{ismail_2021_GLOBECOM}.
With the significant changes at the lower layers of the network to accommodate JRC systems, and due to their multi-objective nature, deriving an optimal channel allocation strategy by the spectrum service provider (SSP) to the JRC nodes becomes more complex and a significant problem of interest. 
In addition, when demands from the JRC nodes exceed the available spectrum, the SSP needs to strategically allocate the available channels to maximize its revenue.}



\textcolor{black}{Several auction mechanisms have been developed to address the problem of overloaded requests for the limited bandwidth in wireless networks~\cite{Zhang_2013_auction_survey, Eraslan_2011, Ahmadi_2014}.
Auction mechanisms are efficient tools to allocate the channels to the users that value them the most.
In such mechanisms, the objective can be to maximize the social welfare of the system which is defined as the sum of all users' utilities.
For instance, the authors in~\cite{ismail_2021_GLOBECOM} designed a simple VCG-like auction mechanism for channel allocation for JRC nodes.
However, traditional auction mechanisms do not consider the uncertainty of the submitted bids of the wireless nodes which can make the derived optimal auction solution infeasible in practice~\cite{BenTal_2009}.}
Here, the bids are derived from the values of channels perceived by the wireless nodes that depend on each device's detection sensitivity and capability.
For instance, when a JRC node calculates the valuation of each channel to derive the bids, it can incorporate some uncertain parameters, e.g., the location of the warden or the channel gain, and therefore its realized utility can be less than expected.
Furthermore, the utilities of some users can become negative and according to prospect theory, users' response to losses is much stronger than that to the corresponding gains~\cite{Amos_1992}. Therefore, a mechanism that considers the uncertainty in bids is needed when deriving the optimal auction solution.



Motivated by the above mentioned issues, in this paper, we develop a robust and efficient multi-item auction mechanism for channel allocation in covert JRC systems under uncertainty of bids. The objective is to develop a risk-averse algorithm for channel allocation that maximizes the social welfare of the system under uncertainty of JRC nodes' valuations.
Different from the traditional Vickrey–Clarke–Groves (VCG) mechanism in which the auctioneer is risk-neutral, in the proposed robust auction mechanism, the auctioneer is risk-averse, i.e., the auctioneer accepts to obtain a lower revenue but with higher confidence, which is practical for several safety critical applications.
To the best of our knowledge, this is the first work that considers the problem of robust channel allocation for covert JRC systems.

The main contributions of our work are as follows:
\begin{enumerate}
    \item We design a novel covert JRC system in which friendly jammers are deployed to transmit artificial noise and prevent wardens from detecting ongoing transmissions of the JRC nodes. The proposed design is shown to enable the JRC nodes to perform covertly their radar sensing and data transmission operations with low detection probability.
    
    \item As the task of designing a reliable channel allocation system by the SSP remains challenging, i.e., immune against uncertainty, we develop a robust multi-item auction mechanism to allocate the channels to the JRC nodes. Unlike previous works on covert communication, we consider the uncertainty about the warden, such as its location, in the design of the auction mechanism and show how the auction outcomes are affected by the uncertainty range of the warden.

    \item The proposed auction based-model guarantees the properties of individual rationality (IR), incentive compatibility (IC), and budget feasibility (BF). This makes the system resilient both against intentional market manipulation attacks and perturbations in the submitted spectrum bids.
    
    \item We conduct extensive simulations to validate the proposed covert JRC system and derive important properties about the proposed robust auction mechanism compared to deterministic auction mechanisms over different scenarios. The JRC nodes are able to covertly perform their radar sensing and data transmission operations while the SSP is able to derive an optimal allocation strategy that reflects its risk-aversion.

\end{enumerate}

The rest of the paper is organized as follows.
Section II reviews related works.
Section III and IV describe our system model and the proposed robust auction mechanism, respectively.
The evaluation results are then presented in section V. Section VI concludes the paper.

\section{Related Works}
Since our work is a novel design that incorporates several techniques, we review related works which study the problem of spectrum allocation in JRC systems, covert communication and robust optimization in wireless networks.


\subsection{Covert Communication}
Different from upper-layer security techniques that try to protect the content from being intercepted by a third party, covert communication has the goal of hiding the ongoing communication itself from wardens.
In~\cite{Bash_2013}, covert communication was first addressed from an information theory perspective where
a square root limit on the number of covertly transmitted bits was derived.
In~\cite{Arghavani_2021}, a zero-sum game was formulated between the transmitter and the jammers as the first player and a set of wardens as the second player. The authors studied how the warden adjusts its detection threshold to increase its detection probability while the transmitter and the jammers vary their transmission power to increase the detection error probability (DEP) at the wardens.
In~\cite{Zheng_2021_covert}, the authors proposed an uncoordinated jammer selection scheme where a set of friendly jammers cooperate to hide an ongoing transmission in a distributed fashion, which is based on their channel gains to the legitimate receiver. However, they only considered the covertness of only one communication channel. If the network has multiple simultaneous transmitters, their proposed jammer selection scheme becomes inefficient as different receivers would have different channel gain thresholds. Furthermore, the additional cost to the transmitter of deploying multiple jammers was not investigated.
Finally, we should note that, to the best of our knowledge, the covertness of JRC systems was not addressed before in the literature.

\subsection{Spectrum Allocation for JRC Systems}

\textcolor{black}{
Spectrum allocation by JRC nodes for both radar sensing and data transmission has received a significant interest in the literature~\cite{Luong_2021_JRC_survey}.
For instance, in~\cite{Scharrenbroich_2016}, a resource management and scheduling (RMS) process is developed to quantify JRC system's performance where different quality-of-service (QoS) metrics are derived for different use cases of JRC nodes. 
In~\cite{Joash_TVT_2021}, authors used deep reinforcement learning to optimally schedule the spectrum resources by JRC nodes.
In~\cite{Shi_2019}, a pricing-based mechanism was proposed for JRC systems but with a focus on handling the interference between the radar and communication functions instead of the market model.
However, less focus was given to the layer where the SSP allocates the spectrum to the JRC nodes.
In~\cite{Nguyen_2021_jrc}, a hierarchical game model was proposed to allocate spectrum resources for OFDM-based JRC systems. A set of SSPs compete with each other to attract more JRC nodes to be their clients while the JRC nodes determine their optimal spectrum demands to maximize their utilities.
In~\cite{ismail_2021_GLOBECOM}, a VCG-like auction mechanism was developed to allocate channels for autonomous vehicles (AVs). However, the authors considered all the channels to have the same valuation, reducing the problem to a single-item auction. 
In addition, existing works on mechanism design for spectrum allocation did not consider the security of the wireless system into their design, e.g., wardens, which can limit the application of the proposed mechanisms in practical scenarios.
Finally, both works in~\cite{Nguyen_2021_jrc} and~\cite{ismail_2021_GLOBECOM} did not consider the uncertainty about users' valuations which can negatively impact the utility of the users. In fact, a quite small perturbation in the submitted bids can make the solution to the auction model infeasible and thus practically meaningless~\cite{BenTal_2009}.
}





\subsection{Robust Optimization in Wireless Networks}
Most of the existing works on optimization problems are based on the assumption that the data defining the constraints and the objective function of the problem are obtained precisely. However, in many practical scenarios, this assumption does not hold and several sources can perturb the obtained data. For instance, wireless networks have different sources of uncertainties that can significantly impact the network performance and valuation metrics such as distance between nodes, CSI and mobile users' dynamic spectrum demands. To tackle this problem, robust optimization has been proposed in which the optimization is performed over an uncertainty set and the objective is to optimize a worst-case function.

Even though robust optimization is being explored for more than two decades~\cite{BenTal_2009, Bertsimas_2004}, very few works adopted robustness in wireless network optimization problems. 
In~\cite{Ye_2008}, the problem of uncertainty about wireless sensors' locations was considered to design a robust solution that maximizes the extracted data, minimizes the consumed energy, and maximizes the network lifetime. The robust solution is defined as the one with the best worst-case objective over the uncertainty set. The authors showed that as the uncertainty set increases in size, the robust solution provides a significant improvement in the worst-case but with the expense of some loss in optimality, known as the price of robustness~\cite{Bertsimas_2004}.
A distributed robust optimization problem was developed in~\cite{Yang_2008} to solve the problem of power and rate control in wireless communication networks under the uncertainty of CSI.
In a recent work~\cite{Dongqing_2020}, a robust optimization approach was considered for mobile data offloading problem, which is formulated as a multi-item auction where the spectrum for mobile users is auctioned by the SSP to offload from the main base station to other access points. The proposed auction mechanism uses historical data from previous bids to determine the winners and payments. However, the proposed model relies only on previous bids and does not consider the realized new bids during the auction process. This makes the derived solution suboptimal as users might change their submitted bids over time.
In addition, a limited discussion was provided about the valuation function of the mobile users, which significantly impacts the derived uncertainty set.
Finally, existing works on mechanism design for spectrum allocation did not consider the security of the wireless system in their design, which can limit the application of the proposed mechanisms in practical scenarios.


Compared with existing works, the proposed auction mechanism in this paper has the following characteristics: (1) we consider the uncertainty in the submitted bids to derive a robust solution that prevents violation of optimal auction properties and is always feasible in practice; (2) we provide examples of relevant uncertainty sets to our problem while the proposed mechanism still works for general uncertainty sets; and (3) we show that our robust mechanism can model the risk level of the SSP towards the solution based on the size of the uncertainty set.
Even though the obtained total social welfare can be lower than the one obtained by deterministic auctions, i.e., which do not consider the uncertainty of bids, it is less risky and more robust in terms of feasibility.

\section{System Model}
In this section, we first describe our proposed covert JRC system and then present the metrics used by the JRC nodes to evaluate the spectrum and derive their bids. Finally, we present the auction market model where we define the utilities of the SSP and the JRC nodes, followed by the properties of the desired optimal auction solution.

\subsection{Covert JRC System}
\begin{figure}[ht!]
    \centering
    \includegraphics[width=.40\textwidth,height=4.5cm]{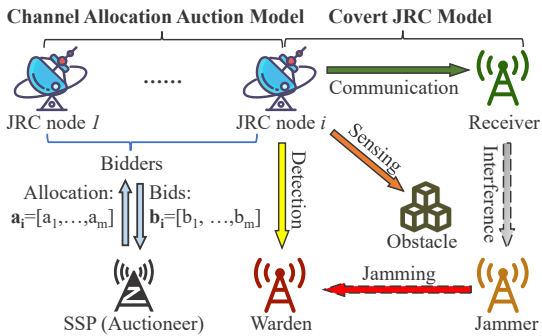}
    \caption{An illustration of the proposed channel allocation auction model and the covert JRC system with the friendly jammer, where a warden is trying to detect the ongoing signals between the JRC node and a receiver/obstacle.}
    \label{fig:covert_comm}
\end{figure}
Figure~\ref{fig:covert_comm} presents the network model under consideration. 
We consider a set $\mathcal{N} = \{1, \dots, N\}$ of JRC nodes. These JRC nodes are under the coverage area of an SSP that has a set $\mathcal{M} = \{1, \dots, M\}$ of channels for allocation, using time-division multiple access (TDMA) to the JRC nodes. 
Therefore, each channel can be used only by one JRC node at a time, which minimizes mutual interference between neighboring JRC nodes.
Each channel has $M_c$ orthogonal sub-carriers that are used by each JRC node to modulate OFDM symbols to transmit simultaneously data and sense the environment~\cite{Liu2017_OFDM, Zhang_2020_MI}.
However, due to the nature of the wireless signal transmission, a warden can detect ongoing data and radar transmissions. To overcome this security issue, we consider that the SSP deploys friendly jammers to lower the probability of warden's success detection to enable the covert JRC system. Friendly jammers help by transmitting random signals to increase the uncertainty at the warden about the ongoing transmission when analyzing the received energy\footnote{This can be further extended to have multiple jammers assisting each JRC node similar to the work presented in~\cite{Zheng_2021_covert}, which adds more uncertainty to the warden about the aggregate interference power.}.
Without loss of generality, we consider that each JRC node is assisted with one jammer and has one warden trying to detect its data transmission and radar sensing. Similar to~\cite{Zheng_2021_covert}, the jamming power is considered fixed for all the jammers to an optimal value that balances between increasing the DEP at the warden and the outage of the legitimate receiver. Note here that the focus of the paper is on channel allocation through the auction mechanism. The optimization of jamming and transmit power of data and radar transmissions can be done, e.g., as in~\cite{Zheng_2021_covert}. The covertness of the radar signals is meant to prevent the warden from identifying the exact JRC node that is trying to sense the environment.

\textcolor{black}{It is interesting to note that on one hand, allocating one jammer for each JRC node might be a waste of resources and one jammer can offer covertness to several JRC nodes. On the other hand, if we use one jammer for more than one JRC node, the jammer will have to transmit at an average power of the covered JRC nodes. This makes some of the JRC nodes much weaker as the used jamming power is far from the optimal jamming power that would be used in the case of single jammer single JRC node scenario, and hence, their transmission can be easily detected by wardens. We should also highlight that existing works, e.g.,~\cite{Soltani_2018_TWC_Covert, Zheng_2021_covert}, suggest that increasing the number of jammers per transmitter brings much more covertness compared to the scenario of single-jammer single-transmitter architecture.
Nevertheless, associating one jammer for each JRC node is an important concern in large scale applications, e.g., device-to-device (D2D) communications for extended coverage. In that scenario, it is possible to use existing JRC nodes as friendly jammers. In this case, how to motivate these nodes to act as friendly jammers would be an interesting issue for further study.}

\subsection{Valuation Metrics}
Different from traditional resource allocation problems, both the radar sensing and data transmission functions need to be jointly optimized in JRC systems. The JRC nodes, therefore, have to consider radar sensing and data transmission performance simultaneously to evaluate the valuation of the spectrum to be acquired through the auction from the SSP. Furthermore, too high transmit power makes the transmitter's sensitive information (e.g., location) more detectable to the warden~\cite{Shahzad_2019}. Therefore, we first analyze the DEP of the warden under generalized fading channels and obtain the closed-form expression of DEP, with arbitrary transmit and jamming power. With the help of the friendly jammer, the transmitter transmits JRC signals while ensuring that the warden's DEP is close to $1$. Only then we can ensure that the communication is covert~\cite{li2020optimal}. With such a precondition, we further analyze the radar and communication performance of the system. We consider the mutual information (MI) between the received signal and the target impulse response to be an important valuation metric for radar systems~\cite{Zhang_2020_MI,Liu2017_OFDM}. The accuracy of the estimated target parameters increases with an increase of the MI~\cite{Liu2017_OFDM}.
In addition, channel capacity (CC) enables the computation of the highest data rate that can be reached via a communication channel and is an important metric for communication systems. 
It has been shown in~\cite{Yang_2007_MI} and~\cite{Liu2017_OFDM} that minimizing the minimum mean square error (MMSE) in estimating the target impulse response is equivalent to maximizing the MI and that careful adjustment of the transmit power according to the channel state information (CSI) increases the data rate.
Therefore, in this work, we adopt MI and CC as the major performance metrics used by the JRC nodes to evaluate the spectrum resources. Because we ensure that the transmission is covert (DEP $\to$ 1), we can call MI and CC as covert MI and covert CC, respectively. 

\subsubsection{Channel Model}
First, we adopt a three-dimensional Cartesian coordinate system to represent locations. The locations of a JRC node $i$, a jammer $g$, a receiver $b$, and a warden $w$ are denoted by ${\mathbf{q}}_i = \left[ {{x_i},{y_i},{z_i}} \right]^T$ , $ {{\mathbf{q}}_g} = {\left[ {{x_g},{y_g},{z_g}} \right]^T}$, ${\mathbf{q}}_b = \left[ x_b,y_b,z_b  \right]^T $ and $ {\mathbf{q}}_w = {\left[ {{x_w},{y_w},{z_w}} \right]^T}$, respectively. The distance between two devices $d_1$ and $d_2$ is expressed as $D_{d_1d_2} ={\|{\mathbf{q}}_{d_1}-{\mathbf{q}}_{d_2}\|}$, and $\alpha_{d_1d_2}$ is the corresponding path loss exponents.
We then use the $\alpha-\mu$ distribution to model the small-scale fading, which is a general fading model that includes several important other distributions, such as the Weibull, One-Sided Gaussian, Rayleigh, and Nakagami. The probability density function (PDF) and the cumulative distribution function (CDF) expressions of a squared $\alpha - \mu$ random variable $\Upsilon$ are given by~\cite{yacoub2007alpha}:
{\small \begin{equation}
	{f_\Upsilon }\left( \gamma  \right) = \frac{{\alpha {\gamma ^{\frac{{\alpha \mu }}{2} - 1}}}}{{2{\beta ^{\frac{{\alpha \mu }}{2} }}\Gamma \left( \mu  \right)}}\exp \left( { - {{\left( {\frac{\gamma }{\beta }} \right)}^{\frac{\alpha }{2}}}} \right),
\end{equation}}\noindent
and
{\small \begin{equation}
{F_\Upsilon }\left( \gamma  \right) = \frac{{\gamma  \left( {\mu ,{\gamma ^{\frac{\alpha }{2}}}{\beta ^{ - \frac{\alpha }{2}}}} \right)}}{{\Gamma \left( \mu  \right)}},
\end{equation}}\noindent
respectively, where {$\Gamma\left(\cdot \right) $} is the gamma function \cite[eq. (8.310.1)]{gradshteyn2007}, {$ \beta  = \frac{{\bar \Upsilon \Gamma \left( \mu  \right)}}{{\Gamma \left( {\mu  + \frac{2}{\alpha }} \right)}} $}, {${\bar \gamma}=E\left(\gamma \right) $}, and $\gamma\left(\cdot \right) $ is the incomplete gamma function \cite[eq. (8.35)]{gradshteyn2007}.
\subsubsection{Detection Error Probability at Warden}
The warden's objective is to minimize the DEP of the ongoing signal transmission, i.e., data and radar signals. For JRC node $i$'s sub-carrier $m$ of channel $j$, the DEP is defined as \cite{zheng2019multi}:
{\small \begin{equation}\label{eq:someequation414514}
{\xi _m^{(ij)}} = \mathcal{P}_{FA} + \mathcal{P}_{MD},
\end{equation}}\noindent
where $\mathcal{P}_{FA}$ is the probability of false alarm, which is defined as $ \Pr \left( {\sigma _c^2 + D_{gw}^{ - {\alpha _{gw}}}{p_g^{(J)}}h_{wg}^2 > {\varepsilon _m}} \right) $, $\mathcal{P}_{MD}$ is the probability of miss detection, which is defined as $ \Pr \left( {D_{iw}^{ - {\alpha _{iw}}}{p_m^{(T)}}h_{wm}^2 + D_{gw}^{ - {\alpha _{gw}}}{p_g^{(J)}}h_{wg}^2 + {\sigma _c^2} < {\varepsilon _m}} \right) $, ${\sigma _c^2}$ is the noise power, ${\varepsilon _m}$ is the detection threshold, $p_m^{(T)}$ is the transmit power, $p_g^{(J)}$ is the jamming power, $ h_{wm}^2 \sim \alpha  - \mu \left( {{\alpha _{wm}^{(ij)}},{\mu _{wm}^{(ij)}},{{\bar \gamma }_{wm}^{(ij)}}} \right) $, and $ h_{wg}^2 \sim \alpha  - \mu \left( {{\alpha _{wg}^{(ij)}},{\mu _{wg}^{(ij)}},{{\bar \gamma }_{wg}^{(ij)}}} \right) $.
As there are $M_c$ sub-carriers for each channel, we consider the DEP for channel $j$ to be the minimum over all DEP for each sub-carrier, i.e.:
{\small \begin{equation}\label{eq:someequation41534454354}
\xi_w^{(ij)}=\min_{m\in M_c}\xi_m^{(ij)} .
\end{equation}}\noindent

\begin{theorem}\label{APPATB}
	The closed-form DEP can be derived as~\eqref{dep}, where $ {C_{1w}} \triangleq D_{iw}^{ - {\alpha _{iw}}}{p_m^{(T)}} $, and $ {C_{2w}} \triangleq D_{gw}^{ - {\alpha _{gw}}}{p_g^{(J)}} $.
\end{theorem}
\begin{IEEEproof}
\renewcommand{\theequation}{A-\arabic{equation}}
\setcounter{equation}{0}
Let $ {Y_1} \triangleq \sigma _c^2 + {C_{2w}}h_{wg}^2 $ and $ {Y_2} \triangleq {C_{1w}}h_{wm}^2 + {C_{2w}}h_{wg}^2 + \sigma _c^2 = {C_{1w}}h_{wm}^2 + {Y_1} $. According to the definition of $ {\xi _m^{(ij)}} $, we have
{\small \begin{equation}\label{defxi}
{\xi _m^{(ij)}} = 1 - {F_{{Y_1}}}\left( {{\varepsilon _m}} \right) + {F_{{Y_2}}}\left( {{\varepsilon _m}} \right).
\end{equation}}

In the following, we derive the CDF expressions of $Y_1$ and $Y_2$. With the help of definition of CDF, we have
{\small \begin{align}
	{F_{{Y_1}}}\!\left( y \right) \!= \!{F_{h_{wg}^2}}\!\!\left(\! {\frac{{y \!- \!\sigma _c^2}}{{{C_{2w}}}}} \!\right) \!= \!\frac{{\gamma\! \left( \!{{\mu _{wg}^{(ij)}},{{\left( {\frac{{y - \sigma _c^2}}{{{C_{2w}}}}} \right)}^{\frac{{{\alpha _{wg}^{(ij)}}}}{2}}}\!{\beta _{wg}^{(ij)}}^{ - \frac{{{\alpha _{wg}^{(ij)}}}}{2}}} \right)}}{{\Gamma \left( {{\mu _{wg}^{(ij)}}} \right)}}.
\end{align}}

The CDF of $Y_2$ can be expressed as \cite{oguntunde2014sum} 
{\small \begin{equation}\label{eqc4}
	{F_{{Y_2}}}\left( y \right) = \int_0^\infty  {{F_{{Y_1}}}\left( {y - t} \right)\frac{1}{{{C_{1w}}}}{f_{h_{wm}^2}}\left( {\frac{t}{{{C_{1w}}}}} \right){\text{d}}t} .
\end{equation}}
Substituting the CDF and PDF expressions in \eqref{eqc4}, with the help of \cite[eq. (06.06.07.0002.01)]{web}, \cite[eq. (01.03.07.0001.01)]{web}, and \cite[eq. (3.194.3)]{gradshteyn2007}, we can express ${F_{{Y_2}}}\left( y \right)$ as \eqref{fy2}, shown at the top of the next page, which can be re-written in closed-form with the help of the definition of multivariate Fox's $H$-function \cite[eq. (A-1)]{mathai2009h}.
\newcounter{mytempeqncntH}
\begin{figure*}[t]
	\normalsize
	\setcounter{mytempeqncntH}{\value{equation}}
	\setcounter{equation}{3}
	{\small \begin{align}\label{fy2}
		&{F_{{Y_2}}}\left( y \right) =\frac{{{\alpha _{wm}^{(ij)}}}}{{2{{\left( {{C_{1w}}{\beta _{wm}^{(ij)}}} \right)}^{\frac{{{\alpha _{wm}^{(ij)}}{\mu _{wm}^{(ij)}}}}{2}}}\Gamma \left( {{\mu _{wm}^{(ij)}}} \right)\Gamma \left( {{\mu _{wg}^{(ij)}}} \right)}}{\left( {y - \sigma _c^2} \right)^{\frac{{{\alpha _{wm}^{(ij)}}{\mu _{wm}^{(ij)}}}}{2}}}{\left( {\frac{1}{{2\pi i}}} \right)^2}
		\notag\\&\times
		\int_{{{\cal L}_1}} {\int_{{{\cal L}_2}} {\frac{{\Gamma\! \left( {1 - \frac{{{s_1}{\alpha _{wg}^{(ij)}}}}{2}} \right)\Gamma\!\left( {{s_1} + {\mu _{wg}^{(ij)}}} \right)\Gamma\!\left( { - {s_1}} \right)\Gamma\!\left( {\frac{{{\alpha _{wm}^{(ij)}}{\mu _{wm}^{(ij)}}}}{2} + \frac{{{s_2}{\alpha _{wm}^{(ij)}}}}{2}} \right)}}{{\Gamma\!\left( {1 + \frac{{{\alpha _{wm}^{(ij)}}{\mu _{wm}^{(ij)}}}}{2} + \frac{{{s_2}{\alpha _{wm}^{(ij)}}}}{2} - \frac{{{s_1}{\alpha _{wg}^{(ij)}}}}{2}} \right)\Gamma\! \left( {1 - {s_1}} \right)\Gamma^{-1}\!\left( { - {s_2}} \right)}}} } 
	{\left( {\frac{{y - \sigma _c^2}}{{{C_{1w}}{\beta _{wm}^{(ij)}}}}} \right)^{\frac{{{s_2}{\alpha _{wm}^{(ij)}}}}{2}}}{\left( {\frac{{y - \sigma _c^2}}{{{C_{2w}}{\beta _{wg}^{(ij)}}}}} \right)^{\frac{{ - {s_1}{\alpha _{wg}^{(ij)}}}}{2}}}d{s_2}d{s_1}
	\end{align}}
	\setcounter{equation}{\value{mytempeqncntH}}
	\hrulefill
\end{figure*}
\setcounter{equation}{4}
Thus, by substituting CDF expressions of $Y_1$ and $Y_2$ into \eqref{defxi}, the DEP can be derived as \eqref{dep}, which completes the proof.
\end{IEEEproof}
	\newcounter{mytempeqncnt}
	\begin{figure*}[t]
		\normalsize
		\setcounter{mytempeqncnt}{\value{equation}}
		\setcounter{equation}{4}
		{\small \begin{align}\label{dep}
			\xi_m^{(ij)}=	&1- \frac{{\gamma \left( {{\mu _{wg}^{(ij)}},{{\left( {\frac{{{\varepsilon _m} - \sigma _c^2}}{{{C_{2w}}}}} \right)}^{\frac{{{\alpha _{wg}^{(ij)}}}}{2}}}{\beta _{wg}^{(ij)}}^{ - \frac{{{\alpha _{wg}^{(ij)}}}}{2}}} \right)}}{{\Gamma \left( {{\mu _{wg}^{(ij)}}} \right)}}
			- \frac{{{\alpha _{wm}^{(ij)}}{{\left( {{\varepsilon _m} - \sigma _c^2} \right)}^{\frac{{{\alpha _{wm}^{(ij)}}{\mu _{wm}^{(ij)}}}}{2}}}}}{{2{{\left( {{C_{1w}}{\beta _{wm}^{(ij)}}} \right)}^{\frac{{{\alpha _{wm}^{(ij)}}{\mu _{wm}^{(ij)}}}}{2}}}\Gamma \left( {{\mu _{wm}^{(ij)}}} \right)\Gamma \left( {{\mu _{wg}^{(ij)}}} \right)}}
			\notag\\&\times
		H_{1,0:1,2:0,0}^{0,1:2,1:1,1}\left(\!\! {\left. {\begin{array}{*{20}{c}}
{\frac{{{C_{1w}}{\beta _{wm}^{(ij)}}}}{{{\varepsilon _m} - \sigma _c^2}}}\\
{\frac{{{C_{2w}}{\beta _{wg}^{(ij)}}}}{{{\varepsilon _m} - \sigma _c^2}}}
\end{array}} \!\!\right|\!\!\begin{array}{*{20}{c}}
{\left( {1 + \frac{{{\alpha _{wm}^{(ij)}}{\mu _{wm}^{(ij)}}}}{2}: - \frac{{{\alpha _{wg}^{(ij)}}}}{2},\frac{{{\alpha _{wm}^{(ij)}}}}{2}} \right):\left( {1 - {\mu _{wg}^{(ij)}},1} \right)\left( {1,1} \right);\left( {1 - \frac{{{\alpha _{wm}^{(ij)}}{\mu _{wm}^{(ij)}}}}{2},\frac{{{\alpha _{wm}^{(ij)}}}}{2}} \right)}\\
{ - :\left( {1, - \frac{{{\alpha _{wg}^{(ij)}}}}{2}} \right)\left( {0,1} \right);\left( {0,1} \right)}
\end{array}} \!\!\right)
		\end{align}}
		\setcounter{equation}{\value{mytempeqncnt}}
		\hrulefill
	\end{figure*}
	\setcounter{equation}{5}

\textcolor{black}{
Note that although the warden's estimate of the channel state is imperfect (including the JRC node's transmit power and the jamming power, which are factors in \eqref{dep}), to verify the robustness of the proposed covert system design, we consider that the warden knows the perfect information (as the worst-case scenario). 
This assumption is actually common in the literature, e.g., in~\cite{Soltani_2018_TWC_Covert, Hongyang_2022_JSAC}.
If we can still guarantee that the DEP is arbitrarily close to 1 under the worst-case scenario, covert communication is successfully achieved. This is also clarified later in the results section (Figure \ref{fig:IJCM}).
}

\subsubsection{Covert Channel Capacity}
The covert CC obtained under the precondition that DEP $\to$ 1, which reflects the covert communication rate. For the channel $j$ of JRC node $i$, the CC is defined as \cite{Liu2017_OFDM,Nguyen_2021_jrc}:
	\setcounter{equation}{5}
{\small \begin{equation}\label{eq:C_i}
{C_{ij}} = \sum\limits_{m = 1}^{{M_c}} \Delta  f{\log _2}\left( {1 + \frac{{D_{bi}^{ - {\alpha _{bi}}}{p_m^{(T)}}{{\left| {{h_m^{(ij)}}} \right|}^2}}}{{\sigma _c^2 + D_{gi}^{ - {\alpha _{gi}}}{p_g^{(J)}}{{\left| {{h_g^{(ij)}}} \right|}^2}}}} \right),
\end{equation}}\noindent
where $ {\left| {{h_m^{(ij)}}} \right|^2} \sim \alpha  - \mu \left( {{\alpha _m^{(ij)}},{\mu _m^{(ij)}},{{\bar \gamma }_m^{(ij)}}} \right) $ and $ {\left| {{h_g^{(ij)}}} \right|^2} \sim \alpha  - \mu \left( {{\alpha _g^{(ij)}},{\mu _g^{(ij)}},{{\bar \gamma }_g^{(ij)}}} \right)$ represent the small scale fading of each sub-carrier $m$ and the jammer, respectively~\cite{yacoub2007alpha}.
${D_{bi}^{ - {\alpha _{bi}}}}$ and ${D_{gi}^{ - {\alpha _{gi}}}}$ denote the large scale fading between the receiver and the JRC node $i$ and between the jammer $j$ and the JRC node $i$, respectively. 
$p_m^{(T)}$ and $p_g^{(J)}$ are the transmit power of the $m$-th sub-carrier and the jammer, respectively. 
$\Delta f = \frac{1}{T}$ is the sub-carrier interval with the duration of elementary OFDM symbol $T$ and $\sigma _c^2$ is the noise variance.
\textcolor{black}{
The jammer's location and its jamming power are publicly shared by the SSP to enable the JRC nodes to calculate the covert channel capacity defined in \eqref{eq:C_i}.
}

\begin{theorem}\label{APPAT}
	The closed-form expression of CC can be derived as \eqref{CFIN}, where $ H_ \cdot ^ \cdot \left(  \cdot  \right) $ is the multivariate Fox's $H$-function \cite[eq. (A-1)]{mathai2009h}, $ {C_1} \triangleq D_{bi}^{ - {\alpha _{bi}}}{p_m^{(T)}} $, and $ {C_2} \triangleq D_{gi}^{ - {\alpha _{gi}}}{p_g^{(J)}} $.
\end{theorem}
	\newcounter{mytempeqncnt3}
	\begin{figure*}[t]
		\normalsize
		\setcounter{mytempeqncnt3}{\value{equation}}
		\setcounter{equation}{6}
		{\small \begin{align}\label{CFIN}
	{C_{ij}} &= \sum\limits_{m = 1}^{{M_c}} \frac{2 \Delta f}{{\ln 2}}\frac{{{{\left( {2{{\left( {{C_2}{\beta _g^{(ij)}}} \right)}^{\frac{{{\alpha _g^{(ij)}}{\mu _g^{(ij)}}}}{2}}}\Gamma \left( {{\mu _g^{(ij)}}} \right)} \right)}^{ - 1}}}}{{{{\left( {{C_1}{\beta _m^{(ij)}}} \right)}^{\frac{{{\alpha _m^{(ij)}}{\mu _m^{(ij)}}}}{2}}}\Gamma \left( {{\mu _m^{(ij)}}} \right)}}{\left( {\sigma _c^2} \right)^{\frac{{{\alpha _m^{(ij)}}{\mu _m^{(ij)}}}}{2} + \frac{{{\alpha _g^{(ij)}}{\mu _g^{(ij)}}}}{2}}}
		\notag\\&\times 
		H_{1,0:3,3;1,1}^{0,1:2,2;1,1}\left(\!\! {\left. {\begin{array}{*{20}{c}}
{\frac{{{C_1}{\beta _m^{(ij)}}}}{{\sigma _c^2}}}\\
{\frac{{{C_2}{\beta _g^{(ij)}}}}{{\sigma _c^2}}}
\end{array}} \!\!\right|\!\!\begin{array}{*{20}{c}}
{\left( {1 + \frac{{{\alpha _m^{(ij)}}{\mu _m^{(ij)}}}}{2} + \frac{{{\alpha _g^{(ij)}}{\mu _g^{(ij)}}}}{2}:1,1} \right):\left( {1,\frac{2}{{{\alpha _m^{(ij)}}}}} \right)\left( {\frac{{{\alpha _m^{(ij)}}{\mu _m^{(ij)}}}}{2},1} \right)\left( {1 \!+\! \frac{{{\alpha _m^{(ij)}}{\mu _m^{(ij)}}}}{2},1} \right);\left( {1,\frac{2}{{{\alpha _g^{(ij)}}}}} \right)}\\
{ - :\left( {\frac{{{\alpha _m^{(ij)}}{\mu _m^{(ij)}}}}{2},1} \right)\left( {\frac{{{\alpha _m^{(ij)}}{\mu _m^{(ij)}}}}{2},1} \right)\left( {1 + \frac{{{\alpha _m^{(ij)}}{\mu _m^{(ij)}}}}{2},1} \right);\left( {\frac{{{\alpha _g^{(ij)}}{\mu _g^{(ij)}}}}{2},1} \right)}
\end{array}} \!\!\!\right)
		\end{align}
		}
		\setcounter{equation}{\value{mytempeqncnt3}}
		\hrulefill
	\end{figure*}
	\setcounter{equation}{7} 
\begin{IEEEproof}
\renewcommand{\theequation}{B-\arabic{equation}}
		\setcounter{equation}{0}
Let $ {C_i} = \sum\limits_{m = 1}^{{M_c}} \Delta  f{C_m} $. The $C_m$ can be expressed as
		{\small \begin{equation}\label{b1}
			{C_m} = \int_0^\infty  {\log \left( {1 + \gamma } \right){f_X}\left( \gamma  \right){\text{d}}\gamma },
		\end{equation}}\noindent
		where $ X \buildrel \Delta \over = \frac{{{C_1}{{\left| {{h_m^{(ij)}}} \right|}^2}}}{{\sigma _c^2 + {C_2}{{\left| {{h_g^{(ij)}}} \right|}^2}}} $. Next, we first derive ${f_X}\left( \gamma  \right)$. Let $ {X_1} = {C_1}{\left| {{h_m^{(ij)}}} \right|^2} $ and $ {X_2} = \sigma _c^2 + {C_2}{\left| {{h_g^{(ij)}}} \right|^2} $, we have ${f_{{X_1}}}\left( x \right) = \frac{1}{{{C_1}}}{f_{{{\left| {{h_m^{(ij)}}} \right|}^2}}}\left( {\frac{x}{{{C_1}}}} \right)$ and ${f_{{X_2}}}\left( x \right) = \frac{1}{{{C_2}}}{f_{{{\left| {{h_g^{(ij)}}} \right|}^2}}}\left( {\frac{{x - \sigma _c^2}}{{{C_2}}}} \right)$. The PDF of $X$ can be expressed as \cite{leonardo2012ratio} $ {f_X}(x) = \int_0^\infty  y {f_{{X_1}}}(xy){f_{{X_2}}}(y){\text{dy}}$. With the
 help of PDF expressions of ${{{X_1}}}$ and ${{X_2}}$, we have 
{\small \begin{equation}
	{f_X}(x)\! = \!\frac{{{x^{\frac{{{\alpha _m^{(ij)}}{\mu _m^{(ij)}}}}{2} - 1}}{{\left(\! {2{{\left(\! {{C_2}{\beta _g^{(ij)}}} \!\right)}^{\frac{{{\alpha _g^{(ij)}}{\mu _g^{(ij)}}}}{2}}}\Gamma \!\left(\! {{\mu _g^{(ij)}}} \right)} \!\right)}^{ - 1}}}}{{2{{\left(\! {{\alpha _g^{(ij)}}{\alpha _m^{(ij)}}} \!\right)}^{ - 1}}{{\left(\! {{C_1}{\beta _m^{(ij)}}} \!\right)}^{\frac{{{\alpha _m^{(ij)}}{\mu _m^{(ij)}}}}{2}}}\Gamma\!\left(\! {{\mu _m^{(ij)}}} \!\right)}}{I_{{A_1}}},
\end{equation}}\noindent
where 
{\small \begin{align}
	{I_{{A_1}}} =& \int_0^\infty  {{{\left( {y - \sigma _c^2} \right)}^{\frac{{{\alpha _g^{(ij)}}{\mu _g^{(ij)}}}}{2} - 1}}\exp \left( { - {{\left( {\frac{{xy}}{{{C_1}{\beta _m^{(ij)}}}}} \right)}^{\frac{{{\alpha _m^{(ij)}}}}{2}}}} \right)} 
	\notag\\&\times
{y^{\frac{{{\alpha _m^{(ij)}}{\mu _m^{(ij)}}}}{2}}}\exp \left( { - {{\left( {\frac{{y - \sigma _c^2}}{{{\beta _g^{(ij)}}{C_2}}}} \right)}^{\frac{{{\alpha _g^{(ij)}}}}{2}}}} \right){\text{dy}}.
\end{align}}

With the help of \cite[eq. (01.03.07.0001.01)]{web}, we can re-write ${f_X}(x)$ as
	{\small \begin{align}\label{A5}
&	{f_X}\!(x)\! =\!\! \frac{{{x^{\frac{{{\alpha _m^{(ij)}}{\mu _m^{(ij)}}}}{2} \!- \!1}}{{\left( {2{{\left(\! {{C_2}{\beta _g^{(ij)}}} \!\right)}^{\frac{{{\alpha _g^{(ij)}}{\mu _g^{(ij)}}}}{2}}}\Gamma\! \left( {{\mu _g^{(ij)}}} \right)} \right)}^{ - \!1}}}}{{2{{\left(\! {{\alpha _g^{(ij)}}{\alpha _m^{(ij)}}} \!\right)}^{ -\! 1}}{{\left( {2\pi i} \right)}^2}{{\left(\! {{C_1}{\beta _m^{(ij)}}}\! \right)}^{\frac{{{\alpha _m^{(ij)}}{\mu _m^{(ij)}}}}{2}}}\Gamma\! \left( {{\mu _m^{(ij)}}} \right)}} \notag\\
	&\times \int_{{\mathcal{L}_1}}{\int_{{\mathcal{L}_2}}\! \!\!{\Gamma \!\left( { -\! {s_1}} \!\right)} }\!{\left( {\frac{1}{{{\beta _g^{(ij)}}{C_2}}}} \right)^{\frac{{{s_2}{\alpha _g^{(ij)}}}}{2}}}{\left( {\frac{x}{{{C_1}{\beta _m^{(ij)}}}}} \right)^{\frac{{{s_1}{\alpha _m^{(ij)}}}}{2}}}\notag\\
	&\times \Gamma \left( { - {s_2}} \right){I_{A1}}d{s_1}d{s_2},
\end{align}}\noindent
where the integration path of $\mathcal{L}_1$ and $\mathcal{L}_2$ goes from $\sigma_1 -i\infty $ to $\sigma_1+i\infty $ and $\sigma_2 -i\infty $ to $\sigma_2+i\infty $, respectively, and $\sigma_1,\sigma_2  \in \mathbb{R}$, ${I_{A1}} = \int_0^\infty  {{y^{\frac{{{\alpha _m^{(ij)}}{\mu _m^{(ij)}}}}{2} + \frac{{{s_1}{\alpha _m^{(ij)}}}}{2}}}{{\left( {y - \sigma _c^2} \right)}^{\frac{{{\alpha _g^{(ij)}}{\mu _g^{(ij)}}}}{2} + \frac{{{s_2}{\alpha _g^{(ij)}}}}{2} - 1}}} {\text{dy}}$, which can be solved with the help of \cite[eq. (3.194.3)]{gradshteyn2007}. Let ${t_1} = \frac{{{s_1}{\alpha _m^{(ij)}}}}{2}$ and $ {t_2} = \frac{{{s_2}{\alpha _g^{(ij)}}}}{2} $. By substituting ${f_X}(x)$ into \eqref{b1}, the $C_m$ can be expressed as \eqref{cmeq}, shown at the top of the next page, where ${I_B} = \int_0^\infty  {\log \left( {1 + \gamma } \right){\gamma ^{{t_1} + \frac{{{\alpha _m^{(ij)}}{\mu _m^{(ij)}}}}{2} - 1}}{\text{d}}\gamma }$.
\newcounter{mytempeqncnt3H}
\begin{figure*}[t]
	\normalsize
	\setcounter{mytempeqncnt3H}{\value{equation}}
	\setcounter{equation}{4}
	{\small \begin{align}\label{cmeq}
			{C_m} =& \frac{4}{{{{\left( {2\pi i} \right)}^2}}}\frac{{{{\left( {2{{\left( {{C_2}{\beta _g^{(ij)}}} \right)}^{\frac{{{\alpha _g^{(ij)}}{\mu _g^{(ij)}}}}{2}}}\Gamma\!\left( {{\mu _g^{(ij)}}} \right)} \right)}^{ - 1}}}}{{2{{\left( {{C_1}{\beta _m^{(ij)}}} \right)}^{\frac{{{\alpha _m^{(ij)}}{\mu _m^{(ij)}}}}{2}}}\Gamma\!\left( {{\mu _m^{(ij)}}} \right)}}{\left( {\sigma _c^2} \right)^{\frac{{{\alpha _m^{(ij)}}{\mu _m^{(ij)}}}}{2} + \frac{{{\alpha _g^{(ij)}}{\mu _g^{(ij)}}}}{2}}}
			\notag\\&\times
			\int_{{{\cal L}_1}} {\int_{{{\cal L}_2}} {\frac{{\Gamma\!\left( { - \frac{{2{t_1}}}{{{\alpha _m^{(ij)}}}}} \right)\Gamma\!\left( {\frac{{{\alpha _g^{(ij)}}{\mu _g^{(ij)}}}}{2} + {t_2}} \right)\Gamma\!\left( { - \frac{{2{t_2}}}{{{\alpha _g^{(ij)}}}}} \right)}}{\Gamma^{-1}\!\left( { - \frac{{{\alpha _m^{(ij)}}{\mu _m^{(ij)}}}}{2} - \frac{{{\alpha _g^{(ij)}}{\mu _g^{(ij)}}}}{2} - {t_1} - {t_2}} \right){\Gamma\!\left( { - \frac{{{\alpha _m^{(ij)}}{\mu _m^{(ij)}}}}{2} - {t_1}} \right)}}} } {\left( {\frac{{\sigma _c^2}}{{{C_1}{\beta _m^{(ij)}}}}} \right)^{{t_1}}}{\left( {\frac{{\sigma _c^2}}{{{C_2}{\beta _g^{(ij)}}}}} \right)^{{t_2}}}{I_B}d{t_1}d{t_2}
	\end{align}}
	\setcounter{equation}{\value{mytempeqncnt3H}}
	\hrulefill
\end{figure*}
\setcounter{equation}{5}
With the help of \cite[eq. (2.6.9.21)]{Prudnikov1986Integrals} and \cite[eq. (8.334.3)]{gradshteyn2007}, $I_{B}$ can be solved.
Substituting $I_{B}$ into \eqref{cmeq}, using the definition of multivariate Fox's $H$-function \cite[eq. (A-1)]{mathai2009h}, we can obtain \eqref{CFIN} to complete the proof.
	\end{IEEEproof}

\textcolor{black}{
Note here that the so called "covert channel capacity" is the same as standard channel capacity with the condition that the DEP is close to 1. If the DEP is far lower than 1, the communication is no longer covert but the capacity of the channel to transmit data is not affected. In other words, the DEP can be less than 0.9 but still the derived channel capacity is correct. However, it is not common in the literature to refer to such system as ``covert``. See for example references~\cite{Zheng_2021_covert, Jiang_2021, Soltani_2018_TWC_Covert}, where the communication is said to be covert only when DEP is between 0.9 and 1.
}

\subsubsection{Covert Radar Mutual Information}
When we ensure that \eqref{dep} is arbitrarily close to $1$ by adjusting the transmit and jamming power, we can guarantee that the signals will not be detected~\cite{li2020optimal}. Now the MI of the JRC system is called \emph{covert radar MI}. For the channel $j$ of JRC node $i$, the MI is defined as~\cite{Liu2017_OFDM,Nguyen_2021_jrc}:
\setcounter{equation}{7}
{\small \begin{align}\label{eq:I_i}
{I_{ij}} = \!\frac{\Delta f{T_{\rm{pri}}}}{2}\sum\limits_{m = 1}^{{M_c}} {{{\log }_2}} \!\!\left(\! {1 \!+\! \frac{{T_{\rm{pri}}D_{bi}^{ - {\alpha _{bi}}}{p_m^{(T)}}{{\left| {G\left( {{f_m}} \right)} \right|}^2}}}{{\Psi\left( {{f_m}} \right) \!+\! D_{gi}^{ - {\alpha _{gi}}}{p_g^{(J)}}{{\left| {J\left( {{f_m}} \right)} \right|}^2}}}} \!\right),
\end{align}}\noindent
where $ {T_{{\rm{pri}}}} = {T_{\rm pulse}}/\delta  $ is the pulse repetition interval of the radar system, $ {T_{\rm pulse}} $ is the radar pulse duration, $ \delta  $ is the radar duty factor, $ {G\left( {{f_m}} \right)}$, $ {J\left( {{f_m}} \right)} $ and $ {\Psi\left( {{f_m}} \right)} $ are energy spectral densities (ESDs) of the transmitted signal, jamming signal and noise, respectively. According to \cite[eq. (5)]{zhang2019mutual}, ESD is viewed as uniform in each sub-channel, and we can consider that $ {\left| {G\left( {{f_m}} \right)} \right|^2} \sim \alpha  - \mu \left( {{\alpha _{rm}^{(ij)}},{\mu _{rm}^{(ij)}},{{\bar \gamma }_{rm}^{(ij)}}} \right) $, $ {\left| {J\left( {{f_m}} \right)} \right|^2} \sim \alpha  - \mu \left( {{\alpha _{rg}^{(ij)}},{\mu _{rg}^{(ij)}},{{\bar \gamma }_{rg}^{(ij)}}} \right) $, and $ {\left| {\Psi\left( {{f_m}} \right)} \right|^2}={\sigma _r^2} $, ${\sigma _r^2}$ is the noise power, $T_{\rm{pri}}$ is the signal duration, and $f_m=f_c+m\Delta f$ is the $m$-th subcarrier frequency with $f_c$ the central frequency. Note that it is possible for the warden to detect reflected signals during radar sensing. However, with the help of the jamming signals, the warden will not be able to know which JRC node has initiated the radar sensing, which is the objective of the covertness for radar sensing.
	
	\begin{theorem}\label{APPAT2}
	The covert radar MI rate can be expressed in closed-form as \eqref{MI}, where $ {C_3} \triangleq {T_{\rm{pri}}}D_{bi}^{ - {\alpha _{bi}}}{p_m^{(T)}} $, and $ {C_4} \triangleq D_{gi}^{ - {\alpha _{gi}}}{p_g^{(J)}} $.
\end{theorem}
	\begin{IEEEproof}
	    Following the similar steps to Theorem \ref{APPAT}, we can derive \eqref{MI} to complete the proof.
	\end{IEEEproof}
	\newcounter{mytempeqncnt2}
	\begin{figure*}[t]
		\normalsize
		\setcounter{mytempeqncnt2}{\value{equation}}
		\setcounter{equation}{8}
		{\small \begin{align}\label{MI}
	{I_{ij}} &= \mathop \sum \limits_{m = 1}^{{M_c}} \frac{{\Delta f{T_{\rm pri}}}}{{\ln 2}}\frac{{{{\left( {2{{\left( {{C_4}{\beta _{rg}^{(ij)}}} \right)}^{\frac{{{\alpha _{rg}^{(ij)}}{\mu _{rg}^{(ij)}}}}{2}}}\Gamma \left( {{\mu _{rg}^{(ij)}}} \right)} \right)}^{ - 1}}}}{{{{\left( {{C_3}{\beta _{rm}^{(ij)}}} \right)}^{\frac{{{\alpha _m^{(ij)}}{\mu _m^{(ij)}}}}{2}}}\Gamma \left( {{\mu _{rm}^{(ij)}}} \right)}}{\left( {\sigma _r^2} \right)^{\frac{{{\alpha _{rm}^{(ij)}}{\mu _{rm}^{(ij)}}}}{2} + \frac{{{\alpha _{rg}^{(ij)}}{\mu _{rg}^{(ij)}}}}{2}}}
		\notag\\&\times H_{1,0:3,3;1,1}^{0,1:2,2;1,1}\!\left(\!\!\! {\left. {\begin{array}{*{20}{c}}
{\frac{{{C_3}{\beta _{rm}^{(ij)}}}}{{\sigma _r^2}}}\\
{\frac{{{C_4}{\beta _{rg}^{(ij)}}}}{{\sigma _r^2}}}
\end{array}} \!\!\!\right|\!\!\!\begin{array}{*{20}{c}}
{\left( {1 \!+ \!\frac{{{\alpha _{rm}^{(ij)}}{\mu _{rm}^{(ij)}}}}{2} + \frac{{{\alpha _{rg}^{(ij)}}{\mu _{rg}^{(ij)}}}}{2}:1,1} \right):\left( {1,\frac{2}{{{\alpha _{rm}^{(ij)}}}}} \right)\left( {\frac{{{\alpha _{rm}^{(ij)}}{\mu _{rm}^{(ij)}}}}{2},1} \right)\left( {1 + \frac{{{\alpha _{rm}^{(ij)}}{\mu _{rm}^{(ij)}}}}{2},1} \right);\left( {1,\frac{2}{{{\alpha _{rg}^{(ij)}}}}} \right)}\\
{ - :\left( {\frac{{{\alpha _{rm}^{(ij)}}{\mu _{rm}^{(ij)}}}}{2},1} \right)\left( {\frac{{{\alpha _{rm}^{(ij)}}{\mu _{rm}^{(ij)}}}}{2},1} \right)\left( {1 + \frac{{{\alpha _{rm}^{(ij)}}{\mu _{rm}^{(ij)}}}}{2},1} \right);\left( {\frac{{{\alpha _{rg}^{(ij)}}{\mu _{rg}^{(ij)}}}}{2},1} \right)}
\end{array}} \!\!\!\right)\end{align}}
		\setcounter{equation}{\value{mytempeqncnt2}}
		\hrulefill
	\end{figure*}
	\setcounter{equation}{9}


\subsection{Auction Model}

Figure~\ref{fig:covert_comm} presents the proposed auction model. We consider that the SSP, as the auctioneer, is offering a unit bundle that consists of a set of channels and friendly jammers to enable covertness for the JRC nodes as the bidders.
The SSP conducts an auction by broadcasting its available spectrum resources to the JRC nodes at every time period $T_b$ (e.g., every 10 seconds). The JRC nodes buy spectrum resources from the SSP and use them for radar sensing and data transmission. 
Each JRC node $i$ submits its bid vector $\mathbf{b}_i=( b_1, b_2, \dots, b_M)$ to the SSP. Each element of the vector $\mathbf{b}_i$ represents the bid that JRC node $i$ is willing to pay for channel $j$. Setting $b_{ij}=0$ means that the JRC node is not interested in channel $j$. 
Before the auction starts, the SSP first calculates the nominal allocation and reservation prices (defined later in Section~\ref{section_auction_based_mech}).
\textcolor{black}{
The calculation of the reservation prices prevents market manipulation by setting a lower bound on acceptable amounts of bids for any JRC node in order to be included in the winner list.}
After receiving the bids from the JRC nodes, the SSP (as the auctioneer) runs the winner selection algorithm to derive the final allocation vector $\mathbf{a}_i=(a_1, a_2, \dots, a_M)$ and the payment ${p_i}$ for each JRC node $i$. 
The winning JRC nodes are then allowed to use the channels according to their allocation vectors $\mathbf{a}_i, \forall i \in \mathcal{N}$. 
In the following, we define the utility functions of the JRC nodes and the SSP and the social welfare maximization problem.


\subsubsection{Utility Functions}
The utility of the SSP is defined as the difference between the payment that it receives from all JRC nodes and the total cost to maintain the channels:
{\small \begin{equation}\label{eq:someequation4332}
u_{SSP} = \sum\limits_{i\in \mathcal{N}} p_{i} - c(\mathbf{x}),
\end{equation}}\noindent
where $p_i$ is the payment given by JRC node $i$ and $c(\mathbf{x}) = \sum\limits_{i\in \mathcal{N}}\sum\limits_{j\in \mathcal{M}} c_j x_{ij}$ is the total channel cost for the allocation vector $\mathbf{x}= \{ x_{ij} \}_{i \in N, j \in M}$ and $c_j$ is the per channel cost for the SSP. The channel cost includes the required computing resources to maintain the channel and the cost of friendly jammers for ensuring the covertness of the JRC system. The cost of channel $j$ is expressed as follows:
{\small \begin{equation}\label{eq:c_j_equation}
\begin{multlined}
    c_j = \kappa_{1,j} p_{\rm{FJ},j} + \kappa_{2,j},
\end{multlined}
\end{equation}}\noindent
where $p_{\rm{FJ},j}$ is the total jamming power used to covert channel $j$, $\kappa_{1,j}$ is the per unit cost of the jamming power, and $\kappa_{2,j}$ is a constant that reflects the licensing fees for channel $j$.

We consider TDMA for the radar and communication functions by the JRC system. Specifically, for some time slots, the allocated channel will be used for radar sensing and then for data transmission in the other time slots. 
Each JRC node $i$ has a private valuation of channel $j$ denoted by $v_{ij}$ which is unknown to the SSP. 
The valuation for each JRC node can vary because of the hardware specific design for each JRC node, e.g., supported wireless technologies that operate on different bandwidths. Also, the valuations given by a JRC node $i$ can differ from one channel to another channel because each channel provided by the SSP can have different transmission characteristics and channel fading parameters.
We define the valuation as follows:
{\small \begin{equation}\label{eq:v_ij_equation}
\begin{multlined}
    v_{ij} = \mathcal{I}_{ij} (\eta_1 I_{ij} + \eta_2 C_{ij})\xi_w,
\end{multlined}
\end{equation}}\noindent

where $\mathcal{I}_{ij}$ is an indicator function in the form of a binary matrix that reflects the ability of JRC node $i$ to use channel $j$ or not, and is known to the SSP. $\eta_1$ and $\eta_2$ are scaling factors, and $\xi_w$ is the DEP at warden $w$. The DEP in (\ref{eq:v_ij_equation}) reflects the discount in the valuation due to the probabilities that the warden detects the ongoing transmission by the JRC node.
\textcolor{black}{
The DEP is chosen to get multiplied into the weighted sum of the covert channel capacity and the covert MI in \eqref{eq:v_ij_equation} because as the DEP decreases, the output of the valuation function in \eqref{eq:v_ij_equation} needs to decrease linearly. If the DEP was just an addition term, the change in the final valuation output would be less apparent\footnote{further mathematical explanation can be found in~\cite{levin_2019_discrete}.}. In other words, in \eqref{eq:v_ij_equation}, we are counting the percentage that we are able to protect against the warden, which is reflected using the DEP value (between 0 and 1). For instance, if the DEP is high (close to 1), this would imply a meaningful allocation to the JRC node, i.e., the performed communication and sensing are successfully covert. Otherwise, if the DEP is low, that indicates a wasted resource allocation.
}

\textcolor{black}{Note that the impact of the two scaling factors cannot be observed beyond the JRC node itself. Specifically, the output of \eqref{eq:v_ij_equation} is just a number which will  be used later during the auction process. Changing the weighting factors will only increase or decrease the submitted bids by the JRC node, i.e., its chances to be among the winners. The form of \eqref{eq:v_ij_equation} has the objective to help the JRC node to determine the best price to submit so as it maximizes its benefit from getting the resource.
The impact of changes of the value computed by \eqref{eq:v_ij_equation} is explored later in the results section.
When the JRC node obtains the spectrum, then based on those coefficients it will allocate the spectrum proportionally for both functionalities based on a TDMA scheme, as discussed in the paper.
Furthermore, the scaling factors can vary dynamically over time based on the JRC node's demand for data transmission or target sensing to assert a certain trade-off as we demonstrated in our previous work~\cite{ismail_2021_GLOBECOM}.}

The JRC node $i$'s utility is then defined as the difference between its valuation for all the channels and its payment $p_i$, which is expressed by the following quasilinear preference function:
{\small \begin{equation}\label{eq:utilityu_JRC_node}
u_i = \begin{cases} 
\sum\limits_{j\in \mathcal{M}} v_{ij}x_{ij} - p_{i}, & \text {if JRC node $i$ wins},
\\ 0, & \text {otherwise}.
\end{cases}
\end{equation}}\noindent

\subsubsection{Social Welfare Maximization}
The solution to the auction mechanism is the maximization of the social welfare function which is defined as the sum of all the utilities, i.e., the utility of the SSP and all the utilities of the JRC nodes. Formally, the social welfare function is defined as:
{\small \begin{equation*}\label{eq:someequation5555}
SW = u_{SSP} + \sum\limits_{i\in \mathcal{N}} u_{i}
\end{equation*}}\noindent
{\small \begin{equation*}\label{eq:SW_0}
= \sum\limits_{i\in \mathcal{N}}\sum\limits_{j\in \mathcal{M}} v_{ij}x_{ij} - \sum\limits_{i\in \mathcal{N}}\sum\limits_{j\in \mathcal{M}} c_j x_{ij}
\end{equation*}}\noindent
{\small \begin{equation}\label{eq:SW_1}
= \sum\limits_{i\in \mathcal{N}}\sum\limits_{j\in \mathcal{M}} (v_{ij}- c_j) x_{ij}.
\end{equation}}\noindent


\subsubsection{Properties of The Auction Mechanism}
Before solving the maximization problem (\ref{eq:SW_1}), the following properties need to be satisfied for an auction to be optimal and efficient:
\begin{itemize}
    \item \textit{\textbf{Incentive compatibility (IC)}:} The JRC node $i$ has no incentive to submit a false bid as for every other bid $v'$, the obtained utility is lower than the utility the JRC node gets by submitting its true valuation $v$. Formally,
    {\small \begin{equation}\label{eq:IC_property}
        \sum\limits_{j\in \mathcal{M}} v'_{ij}x_{ij} -p^{(v')}_i \leq \sum\limits_{j\in \mathcal{M}} v_{ij}x_{ij} -p^{(v)}_i, \quad \forall i\in \mathcal{N},
    \end{equation}}\noindent
    where $p^{(v)}_i$ and $p^{(v')}_i$ are the obtained payments for the true valuation $v$ and any other valuation $v'$, respectively.
    
    \item \textit{\textbf{Budget feasibility (BF)}:}
    The payment vector is budget feasible. Formally,
    {\small \begin{equation}\label{eq:someequation444}
        p^{(v)}_i \leq B_i, \quad \forall i\in \mathcal{N},
    \end{equation}}\noindent
    where $B_i$ is the the maximum budget for JRC node $i$ and is assumed to be publicly known.
    \textcolor{black}{The property of BF is of crucial importance in multi-item auctions. This is because in real systems, the buyers always have a limited budget that they need not to exceed. For example, in an auction mechanism that does not consider BF, if a bidder is selected amongst the winners for several items but he/she cannot pay for all the items, the solution becomes infeasible. One of the main problems in multi-item auctions is that the bidder does not know in advance how many items he/she will win and hence, its budget needs to be incorporated into the optimization problem.}
    
    
    \item \textit{\textbf{Individual rationality (IR)}:}
    The utilities must be non-negative for all JRC nodes, i.e., $u_i \geq 0, \forall i \in \mathcal{N}$
    
    \item \textit{\textbf{Computational efficiency (CE)}:}
    The proposed solution to the optimization problem should be computed in a polynomial time.
\end{itemize}

Therefore, to derive an optimal auction that satisfies the above mentioned properties, problem (\ref{eq:SW_1}) is rewritten as: 
{\small \begin{subequations}
\label{eq:optz_1}
\begin{align}
\begin{split}
\max\limits_\mathbf{x} SW = \sum\limits_{i\in \mathcal{N}}\sum\limits_{j\in \mathcal{M}} (v_{ij}- c_j) x_{ij} \label{eq:optz_1_MaxA} 
\end{split}\\
\begin{split}
\hspace{0.5cm} s.t. \sum\limits_{j\in \mathcal{M}} v_{ij}x_{ij} \leq B_i, \quad \forall i\in \mathcal{N}, \label{eq:optz_1_MaxB}
\end{split}\\
\begin{split}
\hspace{0.5cm} p^{(v)}_i \leq \sum\limits_{j\in \mathcal{M}} v_{ij}x_{ij} , \quad \forall i\in \mathcal{N}, \label{eq:optz_1_MaxC}
\end{split}\\
\begin{split}
\hspace{0.5cm}
\sum\limits_{j\in \mathcal{M}} v'_{ij}x_{ij} -p^{(v')}_i \leq \sum\limits_{j\in \mathcal{M}} v_{ij}x_{ij} -p^{(v})_i, \quad \forall i\in \mathcal{N},
\label{eq:optz_1_MaxC2}
\end{split}\\
\begin{split}
\hspace{1cm} \sum\limits_{i\in \mathcal{N}} x_{ij} \leq 1, \quad \forall j\in \mathcal{M}, \label{eq:optz_1_D}
\end{split}\\
\begin{split}
\hspace{1cm} \mathbf{x} \geq 0, \label{eq:optz_1_E}
\end{split}
\end{align}
\end{subequations}}\noindent
where $x_{ij}$ is the probability that channel $j$ is allocated to the JRC node $i$, and $c_j$ is the cost for each channel $j$. The constraints (\ref{eq:optz_1_MaxB}), (\ref{eq:optz_1_MaxC}) and (\ref{eq:optz_1_MaxC2}), refers to BF, IR and IC, respectively~\cite{Bandi_2014}.
The constraints (\ref{eq:optz_1_D}) and (\ref{eq:optz_1_E}) ensure that the vector of allocation probabilities sums to 1. Note that channels are indivisible items, i.e., each channel is allocated to only one JRC node at a time. Therefore, we are restricted to integral values for the allocation vector $\mathbf{x}$, which we explain later in Algorithm~\ref{algo:RMCA.b}.


\textcolor{black}{
Finally, the SSP needs to take into consideration the uncertainty in bids when deriving the solution to the auction mechanism.
The SSP is able to construct an uncertainty set for these valuations based on the previously submitted bids by the JRC nodes. 
It can then use the constructed uncertainty set during the channel allocation phase to derive an optimal allocation strategy that reflects its risk-aversion attitude about uncertain parameters in the system, e.g., the warden's location.
The size of the uncertainty set determines the risk-aversion level of the SSP, i.e., how robust we want to be. If the SSP has a high level of risk-aversion, it will consider a large uncertainty set and vice-versa.
This has also been validated by other existing works, e.g., \cite{Anna_2008_Risk}, in which the authors showed that the knowledge of a large number of historical data, i.e., a larger uncertainty set, gives an exhaustive set of scenarios, and guarantees the reliability of the derived solution, i.e., high risk-aversion level.
Furthermore, in~\cite{ElGhaoui_2003}, under the assumption of normal distribution, the authors were able to derive an expression that links the size of the uncertainty set to the risk level.
Therefore, in the following section, we develop a robust multi-item auction mechanism that takes into consideration the uncertainty of bids by all the JRC nodes for each channel.}

\section{Auction-based Mechanism for Channel Allocation}\label{section_auction_based_mech}
In this section, we formulate the multi-item auction based JRC resource allocation as a robust optimization problem. The objective is to maximize the social welfare of the system for all valuations by the JRC nodes in the constructed uncertainty set.
Unlike previous works that consider the network geometry to be overt to all the nodes~\cite{Zheng_2021_covert, Shahzad_2019}, the uncertain parameters are typically not known to the transmitters and the SSP, in which case we consider the location of the warden to be the uncertain parameter, while the other uncertain parameters can be adopted in the auction mechanism. The uncertain parameters significantly impact the channel gain equations and  the spectrum valuation, and hence reduce the expected social welfare and violate optimal auction properties. To overcome these challenges, we develop a robust auction mechanism that considers the uncertainty in the bidders' valuations~\cite{Bandi_2014}.

\subsection{Construction of the Uncertainty Set}\label{section_IV:A}
The SSP can create the uncertainty set for the bids based on the type of information it has access to. In the following, we describe two different ways of creating the uncertainty set $\mathcal {U}$ from which the valuation vectors are derived.

\subsubsection{Interval Uncertainty Set}
\textcolor{black}{
In these settings, the belief of the SSP about the valuations of the JRC nodes is modeled based on the lowest and highest possible valuations for each JRC node for each channel. Specifically, the SSP has geometrical information about the transmitter, receiver, and friendly jammer. However, the exact location of the warden is unknown either to the JRC node or to the SSP, affecting the submitted bids by the JRC nodes for each channel. Therefore, the JRC nodes consider that the warden is located in a cube instead of a point in the three-dimensional Cartesian domain.
Then, the JRC nodes can calculate the smallest and largest possible values of DEP at the warden while varying its location inside the cube.} It is also possible to adopt other forms of uncertainty intervals, such as when the warden is located on a sphere. Note that the calculation of the DEPs can be done by methods such as the particle swarm optimization (PSO)~\cite{pso_survey_2018} algorithm. 
Then they substitutes these values in the valuation function presented in (\ref{eq:v_ij_equation}) and derive the lowest and highest valuations for each JRC node and each channel. 
The uncertainty set for channel $j$ with respect to the JRC node $i$ is then defined as follows:
{\small \begin{equation} 
\mathcal {U}_{ij} = \{\mu_{ij} \pm \varsigma_{ij}\},
\end{equation}}\noindent
where $\mu_{ij}$ is the mean value for the valuation of channel $j$ by the JRC node $i$, and $-\varsigma_{ij}$ and $+\varsigma_{ij}$ reflects the minimum and maximum valuations normalized to zero, respectively.
\textcolor{black}{If the SSP has more than one uncertain parameter, it can adopt more generalized techniques for creating the uncertainty set, e.g., correlated historical data technique, presented in the following.
}

\subsubsection{Correlated Historical Data}
If the uncertainty of the SSP about the bids is not limited to the warden's location, i.e., multiple or unknown factors, the belief of the SSP about the valuations of the JRC nodes can be modeled using historical data of previous bids. Specifically, the uncertainty set for channel $j$ is defined as:
{\small \begin{equation} 
\mathcal {U}_j = \left\lbrace \!\!\!\!
\begin{array}{c|c}
(v_{1j},\ldots,v_{Nj}) &
\begin{array}{c}
v_{ij} = f_j + y_{ij}, \quad \forall i \in \mathcal{N},\\
\underline{F_j} \leq f_j \leq \overline{F_j}, \\
-\vartheta \leq \frac{\sum _{i=1}^{N} y_{ij} - N \cdot \mu _g}{\sqrt{N} \cdot \delta _j} \leq \vartheta ,
\end{array}
\end{array}\!\!\!\!
\right\rbrace,
\end{equation}}\noindent
where $f_j$ is a common factor between valuations that reflects the correlation between valuations, $y_{ij}$ are independent components with mean $\mu_j$ and standard deviation $\delta_j$. $\vartheta$ is the parameter that controls the conservativeness of the historical data. 

The robustness is then incorporated in the original problem \eqref{eq:optz_1} as follows:
{\small \begin{equation}
\label{eq:optz_3_robust}
(\mathbf{z}, \mathbf{x}^*) \!=\! \argmax\limits_{\mathbf{v}\in\mathcal{U}}
\left\lbrace\!\!\!\!
\begin{array}{c}\!
\vspace{0.3cm}\max\limits_\mathbf{x} \sum\limits_{i\in \mathcal{N}}\sum\limits_{j\in \mathcal{M}} (v_{ij}- c_j) x_{ij} \\
\vspace{0.3cm}
\hspace{0.5cm} s.t. \sum\limits_{i\in \mathcal{N}} x_{ij} \leq 1, \quad \forall j\in \mathcal{M},\\
\vspace{0.3cm}\sum\limits_{j\in \mathcal{M}} v_{ij}x_{ij} \leq B_i, \quad \forall i\in \mathcal{N}, \\
\vspace{0.3cm} 
\hspace{0cm} 
\sum\limits_{j\in \mathcal{M}} v_{ij}x_{ij} \leq \sum\limits_{j\in \mathcal{M}} u_{ij}x_{ij},\\ \quad \forall\mathbf{u}\in \mathcal{U}, \forall i\in \mathcal{N}, \\
\hspace{1cm} \mathbf{x} \geq 0,
\end{array}\!\!\!
\right\rbrace,
\end{equation}}\noindent
where $\mathbf{z}= \{ z_{ij} \}_{i \in N, j \in M} $ is the optimal valuation vector and the objective is to maximize the worst-case social welfare over all the possible valuation vectors in the uncertainty set $\mathcal{U}$. By setting $\Bar{u}^i_j = \argmin\limits_{\mathbf{u}\in\mathcal{U}}\sum\limits_{j\in\mathcal{M}}x^*_{ij}u_{ij}, \forall i \in\mathcal{N}$, the problem (\ref{eq:optz_3_robust}) is reformulated as follows:
{\small \begin{equation}
\label{eq:optz_4_robust}
(z, x^*) = \argmax\limits_{\mathbf{v}\in\mathcal{U}}
\left\lbrace
\begin{array}{c}
\vspace{0.3cm}\max\limits_x \sum\limits_{i\in \mathcal{N}}\sum\limits_{j\in \mathcal{M}} (v_{ij}- c_j) x_{ij} \\
\vspace{0.3cm}
\hspace{0.5cm} s.t. \sum\limits_{i\in \mathcal{N}} x_{ij} \leq 1, \quad \forall j\in \mathcal{M}, \\
\vspace{0.3cm} \sum\limits_{j\in \mathcal{M}} v_{ij}x_{ij} \leq B_i, \quad \forall i\in \mathcal{N}, \\
\vspace{0.3cm} 
\hspace{0cm} 
\sum\limits_{j\in \mathcal{M}} v_{ij}x_{ij} \leq \sum\limits_{j\in \mathcal{M}} \Bar{u}^i_j x_{ij}, \quad \forall i\in \mathcal{N}, \\
\hspace{1cm} \mathbf{x} \geq 0.
\end{array}
\right\rbrace.
\end{equation}}

The dual of the inner problem (\ref{eq:optz_4_robust}) is as follows:
{\small \begin{equation}
\label{eq:optz_4_dual}
\begin{array}{c}
\vspace{0.3cm}\min\limits_{\boldsymbol{\omega},\boldsymbol{\phi}, \boldsymbol{\psi}} \sum\limits_{j\in\mathcal{M}}\omega_j + \sum\limits_{i\in\mathcal{N}}\left(\phi_i B_i + \psi_i\sum\limits_{j\in\mathcal{M}}x^*_{ij}\Bar{u}^i_j\right) \\
\vspace{0.3cm}
\hspace{0.3cm} s.t.\quad \omega_j + z_{ij} \phi_i + z_{ij} \psi_i + c_j \geq z_{ij}, \quad \forall i\in\mathcal{N}, \forall j\in\mathcal{M} \\
\vspace{0.3cm}
\hspace{1cm} \phi_i, \psi_i \geq 0, \quad \forall i\in\mathcal{N},\\
\vspace{0.3cm}
\hspace{1cm} \omega_j \geq 0,\quad \forall j\in\mathcal{M},
\end{array}
\end{equation}}\noindent
where $\omega_j$, $\phi_i$ and $\psi_i$ are elements of $\boldsymbol{\omega}$, $\boldsymbol{\phi}$ and $\boldsymbol{\psi}$, respectively, and are the duals corresponding to the first, second, and third constraints in (\ref{eq:optz_4_robust}).


\subsection{Robust Mechanism for Channel Allocation (RMCA)}\label{section_IV:B}
The proposed Robust Mechanism for Channel Allocation is executed in two phases:

\subsubsection{Nominal Allocation and Reservation Price Calculation}\label{sec:Nominal:Allocation}
The first phase of the mechanism is executed offline before the beginning of the auction and is presented in Algorithm~\ref{algo:RMCA.a}. The SSP starts first by constructing the uncertainty set as previously described in Section~\ref{section_IV:A} and then uses it as an input to the algorithm with the budgets of each JRC node. Then problem (\ref{eq:optz_3_robust}), which is a bilinear optimization problem and outputs the worst-case valuation vector $\mathbf{z}$ and the nominal allocation vector $\mathbf{x}^*$, is solved using Generalized Benders Decomposition~\cite{Beran_1997}. The dual of the problem (\ref{eq:optz_3_robust}) is then solved to calculate the reservation prices $\mathbf{r}^*=\{r_{ij}\}_{i\in\mathcal{N},j\in\mathcal{M}}$ in steps \ref{algo_1:step_r_begin}-\ref{algo_1:step_r_end}  of Algorithm~\ref{algo:RMCA.a}. The reservation prices are defined as the minimum bids that should be submitted by each JRC node to be admissible to the winner list. As such, if a JRC node submits a bid lower than its reservation prices, it will not be among the winners.
Note that the rationale behind setting the reservation prices equal to the left term of the first constraint of the dual in~\eqref{eq:optz_4_dual} is as follows. Since $\mathbf{z}$ is the optimal solution of the primal for the worst-case valuation, the price that each JRC node has to pay is equal to that at minimum. Otherwise, the SSP (the auctioneer) will have a negative utility.




\begin{algorithm}[ht!]
\SetAlgoLined
\SetKwInOut{Input}{Input}
\SetKwInOut{Output}{Output}
\Input{Uncertainty set $\mathcal{U}$, and budgets $B_1,\dots , B_N$.}
\Output{Reservation prices $\mathbf{r}^*$, and nominal allocations $\mathbf{x}^*$.}

    \SetKwBlock{Beginn}{beginn}{ende}
    \Begin{
        $(z, x^*) \leftarrow$ Solve problem (\ref{eq:optz_3_robust});\\
        $(\omega^*, \phi^*, \psi^*) \leftarrow$ Solve problem (\ref{eq:optz_4_dual});\\
        // Calculate reservation prices\\
        \ForEach{$i \in \mathcal{N}$}{ \label{algo_1:step_r_begin}
        \ForEach{$j \in \mathcal{M}$}{
            $r^*_{ij} = \omega^*_j + z_{ij} \phi^*_i + z_{ij} \psi^*_i + c_j$;
        }
        }\label{algo_1:step_r_end}
    }
 \caption{RMCA.a}
 \label{algo:RMCA.a}
\end{algorithm}

\subsubsection{Final Allocation and Payment Calculation}
The second phase of the mechanism is executed after the bid vector is realized, i.e., the JRC nodes submit their bids to the SSP, and is presented in Algorithm~\ref{algo:RMCA.b}. First, the adapted allocation vector $y^{(v)}$ is calculated by solving the following problem~\eqref{eq:optz_5_yv}, in which the objective is to maximize the social welfare with consideration of the previously derived reservation prices $\mathbf{r^*}$ and the realized bid vector $\mathbf{v}$:
\begin{subequations}
\label{eq:optz_5_yv}
\begin{align}
\begin{split}
\max\limits_{y^{(v)}} \sum\limits_{i\in \mathcal{N}}\sum\limits_{j\in \mathcal{M}} (v_{ij} -c_j - r^*_{ij}) y^{(v)}_{ij} \label{eq:optz_5_MaxA} 
\end{split}\\
\begin{split}
\hspace{0.3cm} s.t. \sum\limits_{i\in\mathcal{N}} y^{(v)}_{ij}\leq 1-\sum\limits_{i\in\mathcal{N}} x^*_{ij}, \quad \forall j\in \mathcal{M} \label{eq:optz_5_MaxB}
\end{split}\\
\begin{split}
\hspace{0.3cm} \sum\limits_{j\in \mathcal{M}}y^{(v)}_{ij} u_{ij} \leq B_i - \sum\limits_{j\in \mathcal{M}}x^*_{ij} r^*_{ij} + \sum\limits_{j\in \mathcal{M}}x^*_{ij}\psi^*_{i}\Bar{u}^i_j,\\
\quad \forall\mathbf{u}\in \mathcal{U}, \forall i\in \mathcal{N}, \label{eq:optz_5_D}
\end{split}\\
\begin{split}
\hspace{1cm} \mathbf{y}^{(v)} \geq 0. \label{eq:optz_5_E}
\end{split}
\end{align}
\end{subequations}

Then we calculate the adapted allocation $y^{(v)_{-k}}$ which is similar to problem (\ref{eq:optz_5_yv}) with JRC node $k$ removed from the set of bidders:
\begin{subequations}
\label{eq:optz_6_yvk}
\begin{align}
\begin{split}
\max\limits_{y^{(v)_{-k}}} \sum\limits_{i\in \mathcal{N}\backslash\{k\}}\sum\limits_{j\in \mathcal{M}} (v_{ij} -c_j - r^*_{ij}) y^{(v)_{-k}}_{ij} \label{eq:optz_6_MaxA} 
\end{split}\\
\begin{split}
\hspace{0.3cm} s.t. \sum\limits_{i\in\mathcal{N}\backslash\{k\}} y^{(v)_{-k}}_{ij}\leq 1-\sum\limits_{i\in\mathcal{N}} x^*_{ij}, \quad \forall j\in \mathcal{M} \label{eq:optz_6_MaxB}
\end{split}\\
\begin{split}
\hspace{0.3cm} \sum\limits_{j\in \mathcal{M}}y^{(v)_{-k}}_{ij} u_{ij} \leq B_i - \sum\limits_{j\in \mathcal{M}}x^*_{ij} r^*_{ij} ,
\quad \forall\mathbf{u}\in \mathcal{U}, \forall i\in \mathcal{N}\backslash\{k\}, \label{eq:optz_6_D}
\end{split}\\
\begin{split}
\hspace{1cm} \mathbf{y}^{(v)_{-k}} \geq 0. \label{eq:optz_6_E}
\end{split}
\end{align}
\end{subequations}

The payments are then calculated by using a VCG-like method, in which the JRC nodes are charged the lowest amount that they could have bid such that they are in the winner list~\cite{Bandi_2014}.


\begin{algorithm}[ht!]
\SetAlgoLined
\SetKwInOut{Input}{Input}
\SetKwInOut{Output}{Output}
\Input{Realized bid vector $\mathbf{v}=\{v_{ij}\}_{i\in\mathcal{N},j\in\mathcal{M}}$, reservation prices $\mathbf{r}^*=\{r_{ij}\}_{i\in\mathcal{N},j\in\mathcal{M}}$, and nominal allocation  $\mathbf{x}^*=\{x_{ij}\}_{i\in\mathcal{N},j\in\mathcal{M}}$}
\Output{Allocation vector $\mathbf{a}^*=\{a_{ij}\}_{i\in\mathcal{N},j\in\mathcal{M}}$, and payment $\mathbf{p}^*=\{p_{ij}\}_{i\in\mathcal{N},j\in\mathcal{M}}$}

    \SetKwBlock{Beginn}{beginn}{ende}
    \Begin{
        \eIf{$\mathbf{v} \notin \mathcal{U}$}
        {\label{algo_2:step_v_in_U}
            Do not allocate any channel to any JRC node and exit the auction.
        }{
            $y^{(v)} \leftarrow$ Solve problem (\ref{eq:optz_5_yv});\\
            \ForEach{$k \in \mathcal{N}$}{
                $y^{(v)_{-k}} \leftarrow$ Solve problem (\ref{eq:optz_6_yvk});\\
            }
            // Calculate the final allocation vector\\
            \ForEach{$i \in \mathcal{N}$}{
            \ForEach{$j \in \mathcal{M}$}{
                $a_{ij}^* = y_{ij}^v + x_{ij}^*$;
            }
            } 
            // Calculate the payment vector\\
            \ForEach{$k \in \mathcal{N}$}{
            $p_k =\sum\limits_{j\in\mathcal{M}}y^{(v)}_{kj}r^*_{kj}+\sum\limits_{j\in\mathcal{M}}x^*_{kj}r^*_{kj} - \sum\limits_{j\in \mathcal{M}}x^*_{kj}\psi^*_{k}\Bar{u}^k_j + $\\ $\sum\limits_{i\in \mathcal{N}\backslash\{k\}}\sum\limits_{j\in \mathcal{M}} (v_{ij}- r^*_{ij}) y^{(v)_{-k}}_{ij} - \sum\limits_{i\in \mathcal{N}\backslash\{k\}}\sum\limits_{j\in \mathcal{M}} (v_{ij}- r^*_{ij}) y^{(v)}_{ij}, \quad \forall k \in \mathcal{N}$;
            }
            
            Allocate the $j$th channel to the $i$th JRC node with probability $a^*_{ij}$ and charge $p_i/\sum_{j\in\mathcal{M}}a^*_{ij}$ to the $i$th JRC node; \label{RMCA.b:last_step}\label{algo_2:step_random}
        }
        
    }
 \caption{RMCA.b}
 \label{algo:RMCA.b}
\end{algorithm}

Since the channels are indivisible items, we are restricted to binary allocations of the channels to the JRC nodes, i.e., each channel is allocated to only one JRC node at a time. Therefore, step~\ref{RMCA.b:last_step} in Algorithm~\ref{algo:RMCA.b} consists of allocating channels randomly based on the allocation vector $\mathbf{a^*}$.
Moreover, the condition $\mathbf{v} \notin \mathcal{U}$ is necessary as if the realized bid vector $\mathbf{v}$ does not belong to the uncertainty set $\mathcal{U}$. As such, the solution to the auction mechanism will be suboptimal as there might be negative utilities, violating the IR property.

\textcolor{black}{
Note here that both the JRC nodes and the SSP deal with the uncertainty but in different phases of the algorithm (RMCA).
Specifically, the first phase of the algorithm is executed offline, i.e., before the JRC nodes start submitting their realized bids (RMCA.a). At this point, and as mentioned is Section 4.1.1, each JRC node calculates the DEP interval using PSO algorithm. The SSP is considered to have collected these valuation before the beginning of the auction. 
The SSP uses the constructed uncertainty set during the nominal allocation phase (RMCA.a) to derive an optimal allocation strategy that reflects its risk-aversion about the warden's location.
Finally, when the JRC nodes want to submit their realized bids, they draw the DEP value from a uniform distribution in the interval between the minimum and maximum values of the DEP. 
}

\begin{theorem}\label{them_1}
	The proposed RMCA has the properties of individual rationality, incentive compatibility and budget feasibility, all in expectation.
\end{theorem}
\begin{IEEEproof}
    Since the channel cost $c_j$ in the objective function of problem~\eqref{eq:optz_3_robust} is constant, we can consider $v_{ij}-c_j$ as one variable. Therefore, with this change of variable, the proof follows from the one derived in~\cite{Bandi_2014}.
\end{IEEEproof}

\textcolor{black}{
\subsection{Discussion on The IC Property}
The most important property that should be satisfied by auctions is IC. The proof of the IC property is omitted here to avoid overloading the paper. Here, we give the intuition behind the proof.
}

\textcolor{black}{
First, we should note that IC is not impacted directly by the uncertainty of bids. The IC is the property that guarantees that the bidders have no incentive to misreport their bids. This is guaranteed if we can prove that any other submitted bids will not bring additional benefit for the bidder than its true valuation, as formulated in \eqref{eq:IC_property}.
As illustrated in Algorithm~\ref{algo:RMCA.a}, the objective of the first phase of RMCA is to derive the nominal allocation, the worst-case valuation vector and the reservation prices. This Algorithm~\ref{algo:RMCA.a} is executed before the beginning of the auction, i.e., before the JRC nodes submit their bids, and use only the constructed uncertainty set. The construction of the uncertainty set, as detailed in Section~\ref{section_IV:A}, is done by the SSP and hence, the uncertainty set cannot be forged.
The second phase of RMCA, described in Algorithm~\ref{algo:RMCA.b}, is executed after the JRC nodes submit their bids, which might be untruthful. However, a malicious JRC node is aware that in the first phase of RMCA, the reservation prices have been calculated and submitting a bid lower than its associated reservation price will prevent the JRC node from being in the winner list. Even though this rational does not totally prevent the malicious JRC node from misreporting its valuation, it can still help preventing the SSP from getting a negative utility, as discussed in Section~\ref{section_IV:B}. The important instruction that prevents a malicious JRC node from misreporting its valuation/bid, is in step-\ref{algo_2:step_v_in_U} of Algorithm~\ref{algo:RMCA.b}. Specifically, if the submitted bid is outside the uncertainty set, which is used to derive the optimal auction solution, the auction process will be reset again and no channel is allocated to any JRC node. In other words, a malicious JRC node cannot misreport its bid because the IC property defined in \eqref{eq:IC_property} is already integrated as a constraint in the optimization problem~\ref{eq:optz_3_robust}, which shows the capability of robust optimization.
}

\textcolor{black}{
An important point to discuss also is that the IC holds only in expectation, as shown in Theorem~\ref{them_1}. The output of the auction model loses the total optimality when we move from divisible items to indivisible items, which is executed in phase 2, i.e., Algorithm~\ref{algo:RMCA.b}, step-\ref{algo_2:step_random}. This is because in the case of indivisible items we are restricted to looking for integral allocations of the items (the channels) to the buyers (the JRC nodes). Instead of allocating the items proportionally according to the allocation vector $a^*$ (which is optimal), the channels are allocated to users randomly where the allocation vector $a^*$ is regarded as a probability vector. 
To the best of our knowledge, and based on a recent report~\cite{Kolesnikov_2022}, there is no proven auction design for multi-item multi-buyer scenario that guarantees total IC. In addition, the adopted auction design in this paper is the only work that can guarantee IC for divisible items due to the inherited properties of robust optimization~\cite{Bandi_2014}.
When we consider uncertainty in the auction design, the IC will hold in expectation for indivisible items (e.g., channels), which means that in some cases it might not hold, but the solution is feasible and solves the problem. Moreover, even if a malicious JRC node knows that the IC might not hold in some cases, it is hard to know exactly under which circumstances. This makes the proposed auction design significantly useful.}

\subsection{Deterministic Mechanism for Channel Allocation}
To evaluate the performance of RMCA, we propose a deterministic mechanism for channel allocation based on RMCA. \textcolor{black}{The deterministic RMCA can be regarded simply as an instance of the original RMCA in which the uncertainty set is considered to contain only the realized bid vector. In other words, by running the deterministic RMCA, the SSP directly derives the solution to the multi-item multi-buyer auction problem without consideration of any perturbation in the submitted bids.
In this settings, the warden is considered to be at a fixed distance from the JRC node and the jammer, and then the DEP is calculated by the JRC node as in~\eqref{eq:v_ij_equation} based on that location\footnote{Later in the experiments section, we show the impact of this assumption caused by not considering the uncertainty in the system.}.}
We first reformulate the inner optimization problem (\ref{eq:optz_3_robust}) by omitting the uncertainty of the valuations, which results in the following linear problem:
{\small \begin{subequations}
\label{eq:optz_7_deterministic}
\begin{align}
\begin{split}
\max\limits_\mathbf{x} \sum\limits_{i\in \mathcal{N}}\sum\limits_{j\in \mathcal{M}} (v_{ij}- c_j) x_{ij} \label{eq:optz_7_MaxA} 
\end{split}\\
\begin{split}
\hspace{0.5cm} s.t. \sum\limits_{i\in \mathcal{N}} x_{ij} \leq 1, \quad \forall j\in \mathcal{M}, \label{eq:optz_7_MaxB}
\end{split}\\
\begin{split}
\hspace{1cm} \sum\limits_{j\in \mathcal{M}} v_{ij}x_{ij} \leq B_i, \quad \forall i\in \mathcal{N}, \label{eq:optz_7_D}
\end{split}\\
\begin{split}
\hspace{1cm} \mathbf{x} \geq 0. \label{eq:optz_7_E}
\end{split}
\end{align}
\end{subequations}}

The dual of problem (\ref{eq:optz_7_deterministic}) is then calculated as follows:
{\small \begin{equation}
\label{eq:optz_7_dual}
\begin{array}{c}
\vspace{0.3cm}\min\limits_{\boldsymbol{\omega},\boldsymbol{\phi}} \sum\limits_{j\in\mathcal{M}}\omega_j + \sum\limits_{i\in\mathcal{N}}\phi_i B_i \\
\vspace{0.3cm}
\hspace{0.3cm} s.t.\quad \omega_j + v_{ij} \phi_i + c_j \geq v_{ij}, \quad \forall i\in\mathcal{N}, \forall j\in\mathcal{M} \\
\vspace{0.3cm}
\hspace{1cm} \phi_i \geq 0, \quad \forall i\in\mathcal{N},\\
\vspace{0.3cm}
\hspace{1cm} \omega_j \geq 0, \quad \forall j\in\mathcal{M},
\end{array}
\end{equation}}\noindent
where $\omega_j$ and $\phi_i$ are elements of $\boldsymbol{\omega}$ and $\boldsymbol{\phi}$, respectively, and are the duals corresponding to the first and second constraints in (\ref{eq:optz_7_deterministic}).

To derive the prices, we need to solve the following problem which is a reduced version of problem (\ref{eq:optz_7_deterministic}) where we remove a JRC node $k$ from the set of bidders and calculate the social welfare:

{\small \begin{subequations}
\label{eq:optz_8_deterministic_k}
\begin{align}
\begin{split}
\max\limits_{x^{-k}} \sum\limits_{i\in \mathcal{N}\backslash \{k\}}\sum\limits_{j\in \mathcal{M}} (v_{ij}- c_j) x^{-k}_{ij} \label{eq:optz_8_MaxA} 
\end{split}\\
\begin{split}
\hspace{0.5cm} s.t. \sum\limits_{i\in \mathcal{N}\backslash \{k\}} x^{-k}_{ij} \leq 1, \quad \forall j\in \mathcal{M}, \label{eq:optz_8_MaxB}
\end{split}\\
\begin{split}
\hspace{1cm} \sum\limits_{j\in \mathcal{M}} v_{ij}x^{-k}_{ij} \leq B_i, \quad \forall i\in \mathcal{N}\backslash \{k\}, \label{eq:optz_8_D}
\end{split}\\
\begin{split}
\hspace{1cm} \mathbf{x}^{-k} \geq 0. \label{eq:optz_8_E}
\end{split}
\end{align}
\end{subequations}}

The proposed mechanism is presented in Algorithm~\ref{algo:RMCA.deterministic}.

\begin{algorithm}[ht!]
\SetAlgoLined
\SetKwInOut{Input}{Input}
\SetKwInOut{Output}{Output}
\Input{Realized bid vector $\mathbf{v}=\{v_{ij}\}_{i\in\mathcal{N},j\in\mathcal{M}}$, and budgets $B_1,\dots , B_N$. }
\Output{Allocation vector $\{a_{ij}^*\}_{i\in\mathcal{N},j\in\mathcal{M}}$, and payment $\{p_{ij}^*\}_{i\in\mathcal{N},j\in\mathcal{M}}$}

    \SetKwBlock{Beginn}{beginn}{ende}
    \Begin{
        $x^* \leftarrow$ Solve problem (\ref{eq:optz_7_deterministic});\\
        $(\omega^*, \phi^*) \leftarrow$ Solve problem (\ref{eq:optz_7_dual});\\
        \ForEach{$k \in \mathcal{N}$}{
            $x^{*_{-k}} \leftarrow$ Solve problem (\ref{eq:optz_8_deterministic_k});\\
        }
        // Calculate reservation prices\\
        \ForEach{$i \in \mathcal{N}$}{
        \ForEach{$j \in \mathcal{M}$}{
            $r^*_{ij} = \omega^*_j + v_{ij} \phi^*_i + c_j$;
        }
        }
        
        // Calculate the payment vector\\
            \ForEach{$k \in \mathcal{N}$}{
            $p_k =\sum\limits_{j\in\mathcal{M}}x^*_{kj}r^*_{kj}  + \sum\limits_{i\in \mathcal{N}\backslash\{k\}}\sum\limits_{j\in \mathcal{M}} (v_{ij}- r^*_{ij}) x^{*_{-k}}_{ij}, \quad \forall k \in \mathcal{N}$;
            }
        Allocate the $j$th channel to the $i$th JRC node with probability $x^*_{ij}$ and charge $p_i/\sum_{j\in\mathcal{M}}x^*_{ij}$ to the $i$th JRC node;
    }
 \caption{Deterministic Mechanism for Channel Allocation.}
 \label{algo:RMCA.deterministic}
\end{algorithm}



    
    




\section{Numerical Results}
In this section, we evaluate the proposed auction mechanisms for channel allocation in covert JRC systems. Specifically, we are interested in analyzing the impact of uncertainty about the warden's location on the obtained social welfare for both the robust and the deterministic auction mechanisms.
Again, we use the latter as a benchmark scheme to evaluate the effectiveness of the former.
We also aim to investigate the impact of the number of channels and JRC nodes on social welfare and computation time.
Our solution is implemented using \emph{Gurobi optimizer} and the python library RSOME~\cite{Chen_2020_rsome} for robust optimization.
Experiments are run on a computer with Intel(R) Xeon(R) CPU at 2.20GHz using 13 GB of RAM and operating on Ubuntu 18.04 system.

We consider a square area of 200 m $\times$ 200 m where a set of JRC nodes, friendly jammers and wardens are located randomly under the coverage of the SSP. Channel costs for the SSP are sampled from a normal distribution with mean $2\$$ and variance $1\$$. The budgets for JRC nodes are chosen uniformly from the interval $[1.5\$, 5\$]$.
As alluded before, we consider that every JRC node has one friendly jammer and one dedicated warden.
Table \ref{table:1} lists the other simulation parameters.

\begin{table}[ht!]
\begin{center}
\caption{Simulation parameters}
\begin{tabular}{ ||p{3.0cm}|P{2.0cm}|| }
 \hline
  Parameter & Value \\ 
 \hline\hline
  Frequency & $5.9 \: GHz$ \\ 
 \hline
  Bandwidth & $50 \: MHz$  \\ 
  \hline
  $p_i^{max}, p_g^{max}$ & 10 dBm  \\
 \hline
 Number of sub-carriers & 10  \\ 
 \hline
  Time-Bandwidth Product & 100  \\ 
 \hline
  Radar Duty Factor & 0.01  \\ 
 \hline
\end{tabular}
\label{table:1}
\end{center}
\end{table}

\subsection{Impact of the jamming power on the covert rate}
\begin{figure}[ht!]
    \centering
    \includegraphics[width=.40\textwidth,height=4.5cm]{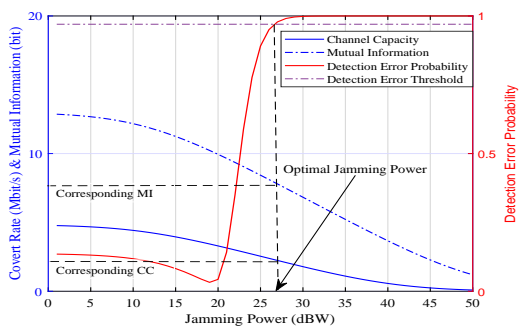}
    \caption{Impact of the jamming power on the channel capacity, mutual information and detection error probability.}
    \label{fig:IJCM}
\end{figure}

To validate our proposed valuation metrics for the covert JRC system, we first conduct Monte Carlo simulations in which we consider a receiver, a JRC node, a friendly jammer and a warden that are located at ${\mathbf{q}}_b = \left[ 7,10,19 \right]$, ${\mathbf{q}}_i = \left[ 3,8,0 \right]$, ${\mathbf{q}}_j = \left[ 6,21,0 \right]$ and ${\mathbf{q}}_w = \left[ 3,14,4 \right]$, respectively.
Figure~\ref{fig:IJCM} depicts the impact of the jamming power on the CC, MI and DEP. 
As we increase the jamming power, the DEP at the warden starts increasing only after the jamming power is greater than $20$ ${\rm dBW}$.
However, we observe that the increase of jamming power causes a decrease of CC and MI. Compared to the case without any jamming signals, to achieve a $97\%$ DEP at the warden, the CC and MI decrease by $50\%$ and $53\%$, respectively. This result suggests that there is a trade-off between the performance and covertness of the JRC system.
Finally, to ensure the covertness of the JRC system, the jamming power must be larger than a certain threshold, i.e., $27$ ${\rm dBW}$ as shown in Figure~\ref{fig:IJCM}. Again, the transmit powers of a JRC node and a friendly jammer can be optimized accordingly, e.g., by using the method provided in~\cite{Zheng_2021_covert}.

\subsection{Uncertainty About Warden's Location}
\begin{figure}[ht!]
    \centering
    \includegraphics[width=.40\textwidth,height=4.5cm]{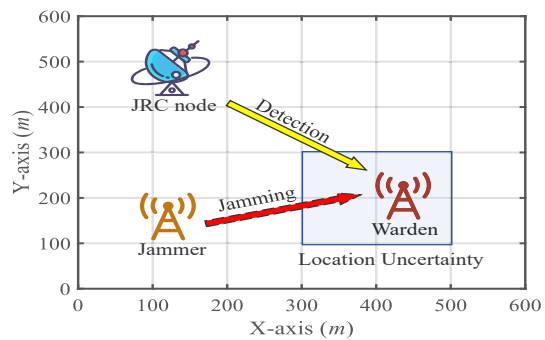}
    \caption{An illustration of uncertainty about warden's location.}
    \label{fig:covert_comm_uncertainty}
\end{figure}

We consider, as an example, the influence of uncertainty about the warden's location on the performance of the system, which is illustrated in Figure~\ref{fig:covert_comm_uncertainty}. Specifically, during the valuation of the spectrum, the JRC node calculates the DEP at the warden based on its belief about the warden's location. However, the JRC node's belief about the warden's location is not accurate and therefore the derived valuations of the JRC nodes might be higher or lower than the real valuations\footnote{Real valuations refer to the derived valuations if the location of the warden is precisely known.}. This implies that the JRC nodes or the SSP might experience negative utilities, violating the IR property of the optimal auction solution.

\begin{figure}[ht!]
    \centering
    \includegraphics[width=.40\textwidth,height=4.5cm]{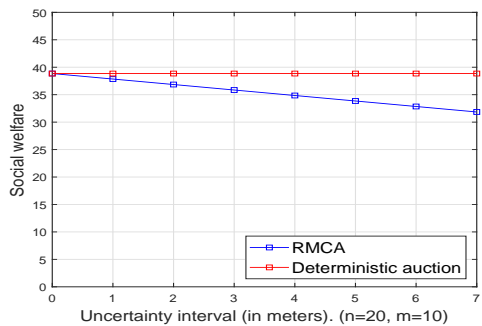}
    \caption{Impact of the uncertainty interval on social welfare.}
    \label{fig:vary_uncertainty_vs_SW}
\end{figure}

To analyze the impact of the uncertainty interval on the social welfare of the system, in this experiment, we set the number of channels to $\mathcal{M}=10$, the number of JRC nodes to $\mathcal{N}=20$, and vary the uncertainty interval about the warden's location, represented in Figure~\ref{fig:covert_comm_uncertainty} by the side of a 2D square surrounding the warden. We observe from Figure~\ref{fig:vary_uncertainty_vs_SW} that the social welfare obtained by the deterministic auction algorithm is not affected by the variations in the uncertainty set, while the social welfare obtained using RMCA decreases as the uncertainty interval,i.e., box width, increases and is lower than that of the deterministic auction algorithm~\cite{Dongqing_2020}. This is explained by the fact that RMCA maximizes the social welfare for the worst-case uncertainty set while the solution derived by the deterministic auction mechanism does not depend on the uncertainty set and uses the realized bids only.


Even though the social welfare obtained by the deterministic auction algorithm is higher than that obtained by RMCA, it comes with a high risk of not being able to be achieved in reality. For instance, if the DEP at warden calculated by the JRC nodes is higher than that if it is in reality, the JRC nodes will have lower utility than expected and can violate the IR property of optimal auctions, i.e., a negative utility. However, the RMCA algorithm is more robust for variations of the DEP at warden which guarantees the feasibility and optimality of the derived solution. The gap between the social welfare obtained by RMCA and the deterministic auction is the price of robustness, i.e., the higher the conservation level about the warden's location, the higher the performance gap between RMCA and the deterministic auction.

\begin{figure}
\centering
\begin{subfigure}{.45\textwidth}
  \centering
  \includegraphics[width=.89\linewidth,height=4.5cm]{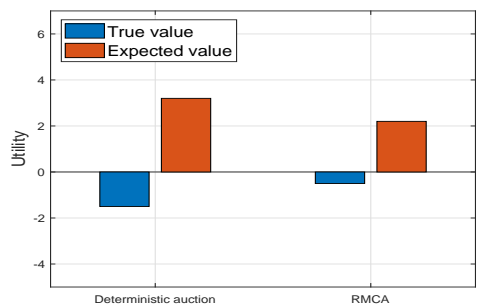}
  \caption{}
  \label{fig:utility_small_range}
\end{subfigure}\\
\begin{subfigure}{.45\textwidth}
  \centering
  \includegraphics[width=.89\linewidth,height=4.5cm]{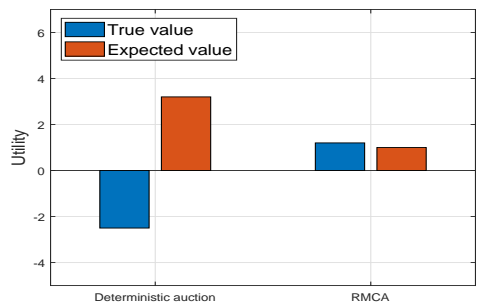}
  \caption{}
  \label{fig:utility_large_range}
\end{subfigure}
\caption{Impact of the uncertainty interval on one of the JRC node's utility in cases where (a) the true location of the warden is outside the uncertainty set, and (b) the true location of the warden is within the uncertainty set. }
\label{fig:utility_JRC_node}
\end{figure}

To further illustrate the robustness of RMCA against the deterministic auction for IR violation, we change the location of the warden to a position where the DEP is lower than expected, i.e., closer to the jammer, then calculate the utility of one of the JRC nodes for both algorithms using \eqref{eq:utilityu_JRC_node}. 
First, we need to distinguish between the \emph{expected utility} and the \emph{true utility}. The expected utility is the utility of a JRC node based on its belief about the warden's location, while the true utility is based on the true location of the warden.
In Figure~\ref{fig:utility_small_range}, we consider that the true location of the warden is outside the uncertainty range defined by the SSP (2 $meters$ in our settings), and in Figure~\ref{fig:utility_large_range} we consider that the warden's location is within the uncertainty range.
From Figure~\ref{fig:utility_JRC_node}, we observe that the expected utility of the deterministic auction is higher than that of RMCA. However, the true utility of the deterministic auction is negative, violating IR. For RMCA, the utility is not negative. In fact, it is negative for RMCA only if the new location of the warden is outside the range from which the uncertainty set is derived. Interestingly, as observed from Figure~\ref{fig:utility_large_range}, when the true location of the warden is within the uncertainty range of the SSP, the true utility derived by RMCA is much higher than the expected utility. This is explained by the fact that RMCA maximizes the worst-case social welfare, i.e., the derived optimal solution is based on the location of the warden that has the lowest DEP.
Note that the knowledge about the violation of the IR property is not possible in real-world scenarios as the location of the warden is usually not known. Therefore, with careful choice of the uncertainty set, the use of RMCA significantly minimizes the chances of violation of the IR property.

\subsection{Impact of mutual information, channel capacity and DEP on the winner list}



\begin{figure}
\centering
\begin{subfigure}{.45\textwidth}
  \centering
  \includegraphics[width=.89\linewidth,height=4.5cm]{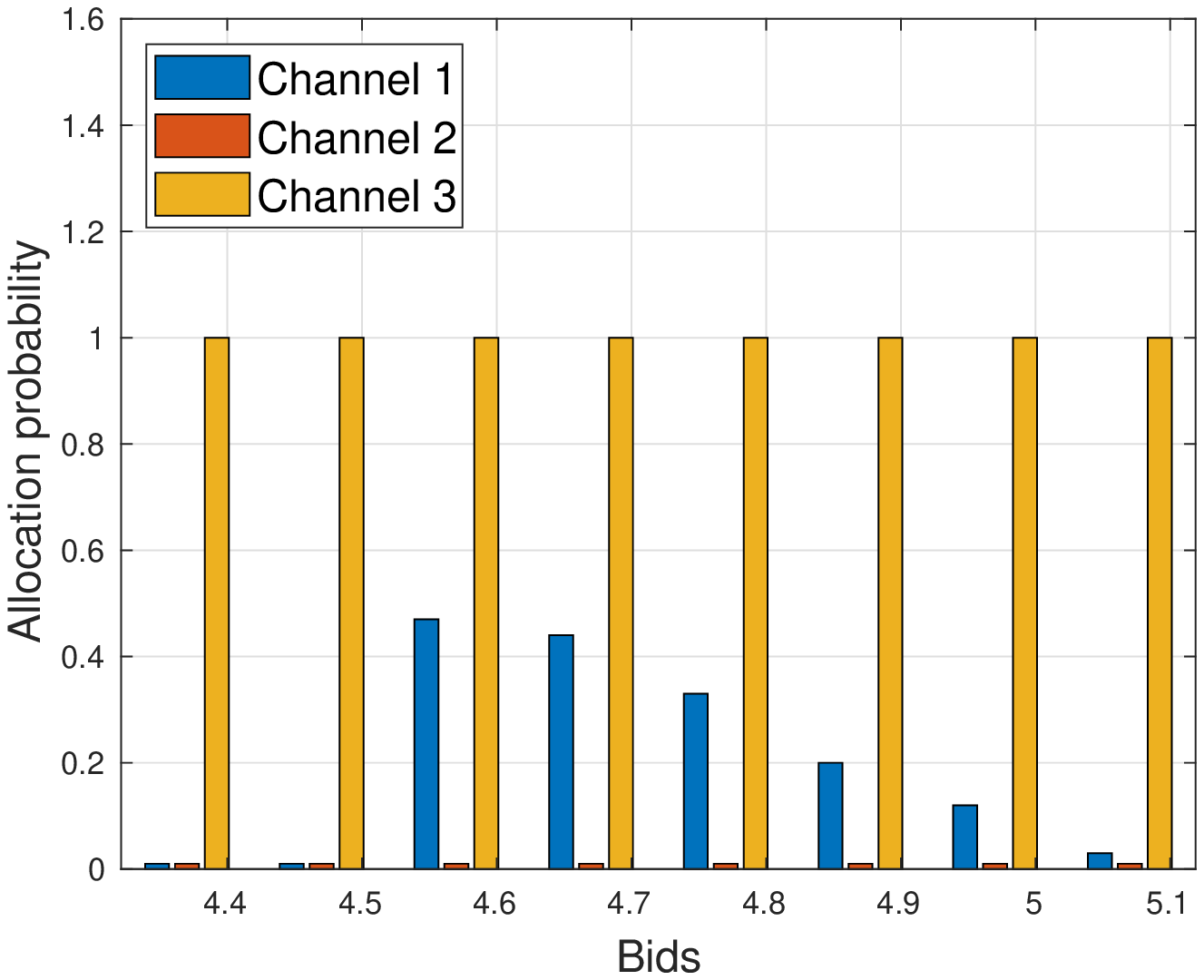}
  \caption{}
  \label{fig:bar_bids}
\end{subfigure}\\
\begin{subfigure}{.45\textwidth}
  \centering
  \includegraphics[width=.89\linewidth,height=4.5cm]{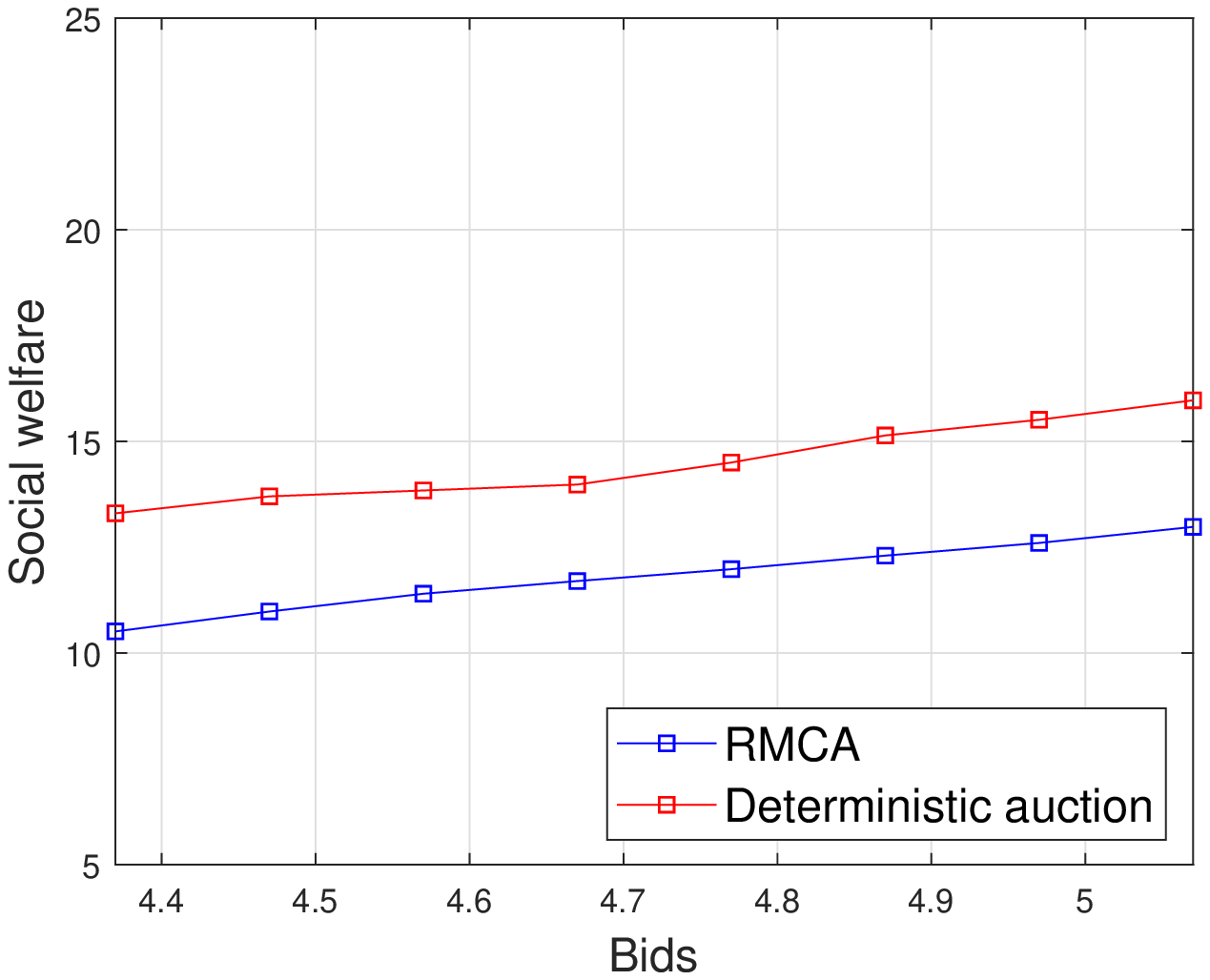}
  \caption{}
  \label{fig:vary_bids_SW}
\end{subfigure}
\caption{Impact of bids of a JRC node on the winner list in terms of (a) the allocation probability and (b) social welfare.}
\label{fig:vary_bids}
\end{figure}


To understand the impact of CC, MI, and DEP at the warden on the winner list, we consider the following scenario.
We set the number of channels to $\mathcal{M}=3$ and the number of JRC nodes to $\mathcal{N}=5$. We then allow one JRC node (ID=5) to change its location so as its valuation for the channels increases based on~\eqref{eq:v_ij_equation}.
Figure~\ref{fig:bar_bids} shows the derived allocation probabilities using RMCA.
We observe that the JRC node 5 is allocated to channel 3 with probability one, and zero for the other channels. However, after its average bids, i.e., the average of the submitted bids by one JRC node to all the channels, increases from 4.47 to 4.57, the allocation probability for JRC node 5 to channel 3 shifts from 0 to 0.45 and then decreases again as the bids increase.
To understand this strange behavior, we show in Table~\ref{table:2}, the submitted bids for all the channels by all the JRC nodes before and after the updated bid values.
We observe that channel 3 is allocated to JRC node 5 with probability 1 because it has a significantly higher bid value than the other JRC nodes for this channel. However, once the bids from the JRC node increase for channel 1 and become the highest, the auction mechanism allocates channel 1 to JRC node 5 with a probability of $45\%$. If we keep increasing the submitted bids of JRC node 5, the allocation probability to channel 1 decreases as shown in Figure~\ref{fig:bar_bids} which is due to the budget constraint, i.e., constraint~\eqref{eq:optz_1_MaxB}. Specifically, since the JRC node 5 is paying more for channel 3, its ability to pay for channel 1 decreases, and hence, the auction mechanism decreases its probability to obtain channel 1.

\begin{table}[ht!]
\begin{center}
\caption{Submitted bids}
\begin{tabular}{ ||P{2.5cm}|P{1.2cm}|P{1.2cm}|P{1.2cm}|| }
 \hline
   & Channel 1 & Channel 2 & Channel 3 \\ 
 \hline\hline
  JRC node 1 & 4.17 & 3.11 & 3.69  \\ 
 \hline
  JRC node 2 & 4.77 & 2.56 & 3.09  \\ 
  \hline
  JRC node 3 & 4.42 & 4.20 & 3.12  \\ 
 \hline
  JRC node 4 & 4.23 & 4.33 & 3.26  \\ 
 \hline
  JRC node 5 & 4.75 & 4.07 & 4.58  \\ 
 \hline \hline
  JRC node 5 (varied) & 4.85 & 4.17 & 4.68  \\ 
 \hline
\end{tabular}
\label{table:2}
\end{center}
\end{table}

We also observe from Figure~\ref{fig:vary_bids_SW} that as the bids from JRC node 5 increase, the social welfare for both algorithms increases. However, the social welfare obtained by RMCA is lower than that obtained by the deterministic one, which is similar to the results shown in Figure~\ref{fig:vary_uncertainty_vs_SW}, i.e., the price of robustness~\cite{Bertsimas_2004}.
Interestingly, our simulations reveal that for the interval-based uncertainty set, there is no difference between RMCA and the deterministic auction algorithm in terms of the allocation probabilities. This is explained by the fact that for our interval-based uncertainty set, RMCA can be regarded as a deterministic auction where the locations of the wardens are fixed at the point with the lowest DEP in the uncertainty box.
Therefore, only social welfare is impacted by the changes in bids but the allocation probabilities are the same for both algorithms.


\subsection{Computation time for different number of JRC nodes and channels}
\begin{figure}[ht!]
    \centering
    \includegraphics[width=.40\textwidth,height=4.5cm]{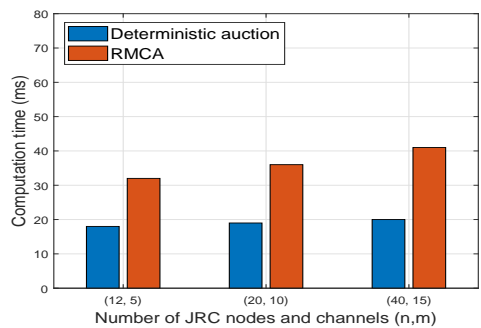}
    \caption{Computation time for different number of JRC nodes and channels.}
    \label{fig:comp_time}
\end{figure}

Figure~\ref{fig:comp_time} shows the computation time for both of the algorithms while varying the number of channels and JRC nodes. We observe that the deterministic auction has almost a constant computation time for different combinations of the number of channels and JRC nodes. However, the RMCA has a higher computation time for the same combinations and increases polynomial with increases in the number of channels and JRC nodes. This is explained by the fact that the RMCA solves a bilinear optimization problem, i.e., problem (\ref{eq:optz_3_robust}), which is NP-hard for general uncertainty sets $\mathcal{U}$. However, the computation time is still tractable which is due to the use of the Generalized Benders Decomposition algorithm that assures a polynomial computation time if the uncertainty set has a polynomial number of extreme points which is the case in our interval-based uncertainty set $\mathcal{U}$.
Note that since the first phase of RMCA, i.e., RMCA.a, is executed before bid submission, the computation time can be further reduced if the set of participating JRC nodes is the same as in the previous round of the auction.

\subsection{Discussions}
From the results obtained from Figure~\ref{fig:vary_bids}, we observe that even though a JRC node increases its bid, its chances of getting a channel decrease because of the budget constraint. In this case, the SSP obtains much higher revenue but the JRC node's utility decreases with no additional benefit. In other words, the JRC node can be allocated to the same number of channels if it bids untruthfully, violating the IC property. This case occurs because the developed mechanism, as earlier shown, is only IC in expectation and not dominant strategy incentive compatible (DSIC). However, obtaining the minimum bid value by a JRC node is practically difficult because the mechanism is probabilistic and the JRC node's objective is to maximize its chances of getting the channel. Moreover, JRC nodes are not aware of other bids. Therefore, the JRC nodes are incentivized to bid truthfully.

\textcolor{black}{
Another main observation from our results is the distinction between ex-ante IR and ex-poste IR.
Ex-ante IR refers to the case where the JRC node anticipates that it has a non-negative expected utility before
the winner list and prices are determined, while ex-post IR refers to the case where the JRC node is guaranteed to have a non-negative utility after the winner list and prices are determined. 
The ex-post IR property is certainly desired in our system. This is because it represents the true utility that the JRC nodes get. Based on~\eqref{eq:utilityu_JRC_node}, getting a negative utility implies that the JRC node is paying more that it gets in benefits, which makes the JRC node reluctant to participate in such auctions in the future.
As earlier shown in Figure~\ref{fig:utility_JRC_node}, the deterministic RMCA has ex-ante IR but lacks ex-post IR. However, RMCA can guarantee ex-post IR if the uncertainty set is well defined, i.e., large enough to include all possible realization of the bids. The deterministic RMCA cannot have this guarantee even with large uncertainty sets. 
}

\textcolor{black}{Finally, the developed deterministic auction mechanism opens an interesting research area to explore. Specifically, the use of robust optimization tools has enabled us to derive optimal solutions that have the properties of IC, IR and BF. The power of robust optimization is that these properties are smoothly incorporated in the optimization problem, making the solution to the auction problem significantly easier than existing complicated auction designs. This suggests that we can use robust optimization for other auction problems where we need to guarantee the properties of IC, IR and BF and then, we might omit the discussion about uncertainty by simply considering the uncertainty set to contain only one item (as we have done in our deterministic auction). Certainly, these suggestions needs further investigations and validation as there might appear other challenges.}

\section{Conclusion}
In this paper, a covert JRC system that can operate safely in the existence of a watchful adversary has been developed. The reliability of the channel allocation problem by the SSP to the JRC nodes was addressed, where we proposed a robust auction mechanism to maximize the social welfare of the system. The proposed auction mechanism was shown to be robust against perturbations in the submitted bids. We implemented a deterministic auction mechanism to show the benefits of robustness. Simulation results showed that the robust auction mechanism yields better performance compared to the deterministic auction mechanism in terms of satisfaction of the optimal auction solution when there is uncertainty about the submitted bids. For future works, we would like to investigate the use of deep learning in auction design and study what type of robustness it can provide.

\section{Acknowledgement}
The authors would like to thank Prof. Chaithanya Bandi for the valuable suggestions that helped us significantly improve the paper.



    



\bibliographystyle{IEEEtran}
\bibliography{ref3}

\end{document}